\documentclass[aps,prb,twocolumn,superscriptaddress,floatfix]{revtex4-2}

\usepackage[utf8]{inputenc}
\usepackage[english]{babel}
\usepackage{mathtools}
\usepackage{physics}
\usepackage{float}
\usepackage{braket}
\usepackage{bbm}
\usepackage{bm}
\usepackage{amsbsy}
\usepackage{amsthm}
\usepackage{amssymb}
\usepackage{amsfonts}
\usepackage{amsmath}
\usepackage{dsfont} 
\usepackage{graphicx} 
\usepackage{hhline}
\usepackage{epsfig}
\usepackage{epstopdf}
\usepackage{dsfont}
\usepackage{multibib}
\usepackage{xcolor}
\usepackage{color}
\usepackage{verbatim}
\usepackage{nomencl}
\usepackage[colorlinks]{hyperref}
\usepackage{float}
\usepackage[normalem]{ulem}
\usepackage{multirow}
\usepackage{makecell}

\begin{document}
	
	\title{Sufficient condition for gapless spin-boson Lindbladians, and its connection to dissipative time-crystals}
	
	\author{Leonardo da Silva Souza}
	\affiliation{Instituto de F\'isica, Universidade Federal Fluminense, Av. Gal. Milton Tavares de Souza s/n, Gragoat\'a, 24210-346 Niter\'oi, Rio de Janeiro, Brazil}
	\author{ Luis Fernando dos Prazeres}
	\affiliation{Instituto de F\'isica, Universidade Federal Fluminense, Av. Gal. Milton Tavares de Souza s/n, Gragoat\'a, 24210-346 Niter\'oi, Rio de Janeiro, Brazil}
	\author{ Fernando Iemini}
	\affiliation{Instituto de F\'isica, Universidade Federal Fluminense, Av. Gal. Milton Tavares de Souza s/n, Gragoat\'a, 24210-346 Niter\'oi, Rio de Janeiro, Brazil}
	
	\begin{abstract}
		We discuss a sufficient condition for gapless excitations in the Lindbladian master equation for collective spin-boson systems and permutationally invariant systems. The condition relates a nonzero macroscopic cumulant correlation in the steady state to the presence of gapless modes in the Lindbladian. In phases arising from competing coherent and dissipative Lindbladian terms, we argue that such gapless modes, concomitant with angular momentum conservation, can lead to persistent dynamics in the spin observables with the possible formation of dissipative time-crystals. 
		We study different models within this perspective, from Lindbladians with Hermitian jump operators, to non-Hermitian ones composed by collective spins and Floquet spin-boson systems.
		We also provide a simple analytical proof for the exactness of mean-field semiclassical approach in such systems based on a cumulant expansion. 
	\end{abstract}

	\maketitle
	
	
	Nonequilibrium quantum many-body dynamics constitutes a fundamental and open research field~\cite{alessio206,Abanin2019,Maldacena2016}. The dissipative dynamics of a quantum system embedded in a environment can, in general, be quite cumbersome due to its high complexity. A common useful approach relies on Born-Markovian approximation, for which the effective dynamics for the quantum system is described by a Lindbladian master equation~\cite{petruccione}. Among its emergent phases, a new form of spontaneous symmetry breaking (SSB) so-called dissipative time-crystal~\cite{Wilczek2012} has gained much attention recently. These nonequilibrium phases break spontaneously the time-translation symmetry of the system, leading to persistent oscillations of macroscopic observables in the thermodynamic limit. Despite intense theoretical and experimental activity (see \cite{Sacha2018,Dominic2020} for interesting reviews) many aspects of these new phases  are still being unraveled, with particular attention to the precise role of its many-body correlations~\cite{Antonio2022,carollo2022praletter},  symmetries~\cite{Giulia2021,Luis2021,Yuma2022,Gianluca2022} and  basic mechanisms for the stabilization of such peculiar nonequilibrium phases.
	
	The spectral properties of a Lindbladian master equation host valuable information on the system dynamics and phases~\cite{biella2018,Albert2014,prozen2008,Iemini2018,Gong2018,Berislav2019}. The Lindbladian gap in particular can characterize the critical behavior in dissipative phase transitions, the emergence of symmetry breaking phases as well as the asymptotic relaxation dynamics towards the steady states of the system. 
	In dissipative TC's  the Lindbladian spectrum features gapless excitations 
	generating long-lived asymptotic dynamics towards the steady state, with a divergent lifetime in the thermodynamic limit~\cite{Iemini2018,Gong2018}. 
	These gapless excitations appear along with coherent dynamics within their subspace inducing the persistent oscillations of the system observables.
	The determination of the Lindbladian gap, however, is not in general a trivial task. Apart from models sharing specific structures~\cite{ribeiro2019,McDonald2022,Aleksandra2020,Nakagawa2021} (as quadratic fermion/boson Lindbladians, symmetries, integrability) for which one can determine it spectral properties and steady states analytically (or quasi-analytically), for general interacting systems its computation relies either in the diagonalization of the Lindbladian superoperator in an extended Hilbert space or inferring from the dynamics of the observables of the system in the asymptotic limit; in both cases an often nontrivial and challenging task.

\begin{figure}
		\includegraphics[width = 0.8 \linewidth]{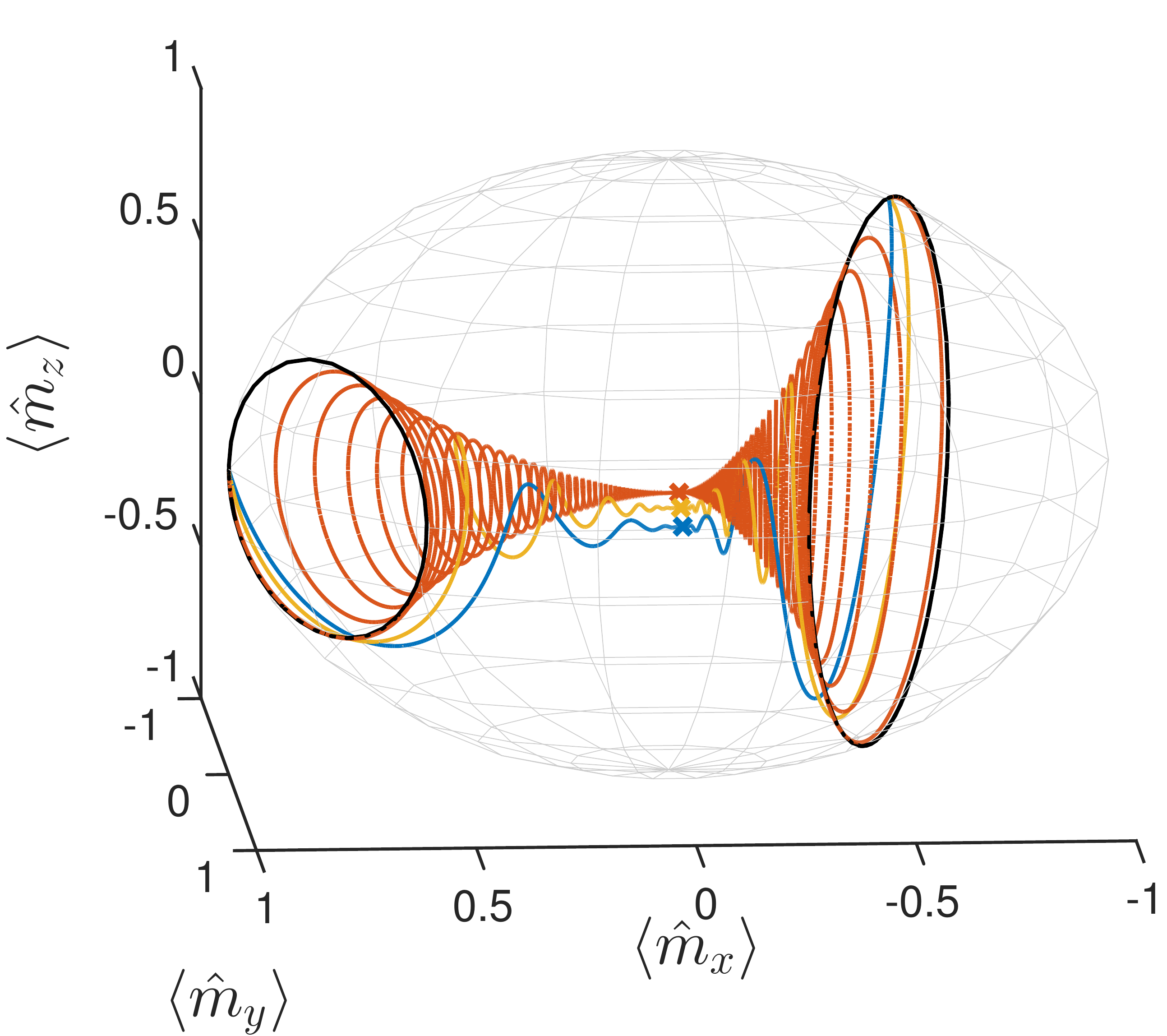}
		\caption{ Dynamics of macroscopic spin magnetizations $\langle \hat m^\alpha \rangle$ for a collective spin-$1/2$ model with competing coherent Hamiltonian ($\omega_x$) and dissipation $\kappa$ (Eq.\eqref{eq:dissipation.su2}), for $\omega_x/\kappa = 2$. We show the  dynamics for different initial conditions and system sizes - $N=2^2$ (blue), $2^3$ (yellow), $2^6$ (red) and semiclassical limit $N\rightarrow \infty$ (black). The crosses are the corresponding NESS, and the spherical shell represents the semiclassical set of states with null macroscopic cumulants, $\langle \hat m^x \rangle^2 +\langle \hat m^y \rangle^2 + \langle \hat m^z \rangle^2 = 1 $. The NESS lie inside the spherical shell (i.e., with nonzero macroscopic cumulant correlations). In the thermodynamic limit the dynamics is constrained to the shell and therefore cannot reach such NESS leading to the appearance of gapless Lindbladian excitations, featuring in this case a persistent dynamics.
		}
		\label{fig:shell.su2}
	\end{figure}

	In this work we discuss a simple sufficient condition to ensure the gapless nature of a Lindbladian, based only on its steady state correlations. Although (long-range) correlations are expected to be connected to gapless excitations and ground state SSB phases in closed Hamiltonian settings, in this work we obtain an analytical proof of this correspondence for a class of open-systems driven by Lindbladian master equation. The condition is based on the exactness of mean-field semiclassical approach for such systems, which we proof using a cumulant expansion, and the fact that the macroscopic spin magnetizations of the nonequilibrium steady state (NESS) cannot be dynamically reached in the thermodynamic limit due to spin total angular momentum conservation (as illustrated in Fig.\eqref{fig:shell.su2}). Studying different collective spin-boson models we see that such gapless excitations, or the inability of the dynamics to reproduce the spin NESS magnetizations, are usually associated to the appearance of persistent dynamics and possible dissipative TC phase, indicating an intimate connection among them.
	
	\textbf{\textit{The model.- } } We consider $M$ spin ensembles, each composed of $N$ spin-$1/2$ subsystems, interacting with a bosonic mode and coupled to a Markovian environment. The time evolution for the system is described by the Lindbladian master equation~\cite{petruccione},
	\begin{equation}\label{eq.master.equation.spin}
		\frac{d}{dt} \hat\rho = \mathcal{ L}[\hat \rho] =  -i [\hat H,\hat\rho ] +\sum_{i=1}^M \mathcal{D}_i\left[\rho\right] + \mathcal{D}_a\left[\rho\right],
	\end{equation}
	with $\mathcal{L}$ the Lindbladian superoperator. The coherent driving term is given by 
	$ \hat H = \hat H_{\rm spin} + \hat H_{\rm boson} +  \hat H_{\rm spin-boson},$ 
	where $\hat H_{\rm spin (boson)}$  corresponds to the spin (boson) term and $\hat H_{\rm spin-boson}$ to the spin-boson interaction. Specifically,
	\begin{eqnarray}\label{eq.Ham.definition.eq1}
		\hat H_{\rm spin} &=& \sum_{i=1}^M \sum_{\alpha}\omega_{\alpha}^{(i)}\hat S_{i}^{\alpha} +  \frac{1}{S}\sum_{i,j=1}^M \sum_{\alpha,\beta} \omega_{\alpha,\beta}^{(i,j)}\hat S_{i}^\alpha\hat S_{j}^\beta, \nonumber  \\
		\hat H_{\rm boson} &=& \omega_b \hat a^\dagger \hat a, \\
		\hat H_{\rm spin-boson} &=& \sum_{i=1}^M  \sum_{\alpha}   \frac{g_{\alpha}}{\sqrt{N}}\left(
		\hat{a}+\hat{a}^{\dagger}\right)\hat S^\alpha_i
		, \nonumber 
	\end{eqnarray}
	where $\hat S_{i}^\alpha =  \sum_{k=1}^N \hat \sigma_{i,k}^\alpha/2$~\cite{footnote0}, with $\alpha = x,y,z$ are the collective spin operators for the $i$'th ensemble, $\hat \sigma_{i,k}^\alpha$ are the Pauli spin operators for the $k$'th spin in the $i$'th ensemble, $S=N/2$ is the total angular momentum of the system and $\hat a$ the annihilation operator of the bosonic mode satisfying  $[\hat a, \hat a^\dagger] = 1$. 
    The parameter $w_\alpha^{(i)}$ describe local fields on the collective spins, $w_{\alpha,\beta}^{(i,j)}$  the collective spin-spin interactions, $\omega_b$ the bosonic mode excitation energy and $g_\alpha$ the spin-boson couplings.
	The collective spin operators inherit the $SU(2)$ algebra of their subsystems satisfying the commutation relations $[\hat S_{i}^\alpha,\hat S_{j}^\beta] = i\epsilon^{\alpha \beta \gamma}\hat S_{i}^\gamma\delta_{i,j}$. Due to the collective nature of the interactions, the model conserves the total spin $S_i^2 = (\hat S_{i}^x)^2 +  (\hat S_{i}^y)^2 + (\hat S_{i}^z)^2$.
	
	The dissipative boson and spin terms of the Lindbladians are defined by,
	\begin{eqnarray}\label{eq.dissipation.boson}
		\mathcal{D}_a \left[\rho\right]&=& \kappa_b  \left(\hat a \hat\rho \hat a^\dagger - \frac{1}{2}\{ \hat a^{\dagger} \hat a,\hat \rho \}\right), \\
		\mathcal{D}_i\left[\rho\right]&=&\frac{1}{S}\sum_{\alpha,\beta}\Gamma_{\alpha,\beta}^{(i)} \left(\hat S_{i}^{\alpha} \hat\rho  \hat S_{i}^{\beta} - \frac{1}{2}\{\hat S_{i}^{\beta} \hat S_{i}^{\alpha},\hat\rho \}\right), 
		\label{eq.dissipation.spin}
	\end{eqnarray}
	with $\kappa_b$ representing the boson loss rate and $\Gamma_{\alpha,\beta}^{(i)}\in\mathbb{C}$ the elements of the dissipative spin matrix $\mathbf{\Gamma^{(i)}}\in\mathbb{C}^{3\text{x}3}$ with $\alpha,\beta = x,y,z$. Although in the examples discussed in the manuscript we consider cases with $\mathbf{\Gamma^{(i)}}$ positive semidefinite, representing in fact a Lindbladian master equation, our proofs are independent of such constraint, thus valid for more general dynamics not fully in Lindblad form.
	
	\textbf{\textit{Exactness of Mean-Field (MF) Approach.- }} Due to the collective character of the spin operators, the spins within each ensemble are permutationally invariant. This symmetry simplify considerably the description of the  macroscopic observables 
	\begin{equation}
		\hat x =\frac{\hat{a}^{\dagger} + \hat a }{\sqrt{2N\omega_b}},\quad \hat p =i\frac{\hat{a}^{\dagger}-\hat a}{\sqrt{2N/\omega_b}},\quad \hat{m}_i^{\alpha}=\hat{S}_i^{\alpha}/S,
	\end{equation}
	in the thermodynamic limit. Specifically, in this limit MF is proven exact as we show using a cumulant expansion approach~\cite{SM,Gardiner1985,Kubo1962,Wunsche2015} (different proofs are also known~\cite{Benatti2018,Carollo2021}  using different methods). Such approach shall be useful both to (i) define variables of interest for our gapless condition, (ii) set clear limits of validation for the MF; and (iii) extend the proof for more general systems (discussed later in the manuscript). The reasoning behind our proof is as follows. The first, second and third order cumulants of general observables are defined, respectively, by
	\begin{eqnarray}
		\label{eq.123.cumulant}
		& K(\hat O_j) = \langle \hat O_j \rangle, \quad  
		K(\hat O_j,\hat O_\ell) = \langle \hat O_j \hat O_\ell \rangle - \langle\hat O_j\rangle\langle\hat O_\ell\rangle, & \nonumber \\
		&K(\hat O_j,\hat O_\ell, \hat O_m) = \langle \hat O_j \hat O_\ell \hat O_m\rangle +2\langle \hat O_j\rangle\langle \hat O_\ell\rangle\langle \hat O_m\rangle + \nonumber \\
		&\qquad-\langle \hat O_j \hat O_\ell\rangle\langle\hat O_m\rangle-\langle \hat O_j \hat O_m\rangle\langle\hat O_\ell\rangle -\langle\hat O_\ell \hat O_m\rangle\langle\hat O_j\rangle, 
	\end{eqnarray}
	Deriving the Heisenberg equations of motion for the second cumulant of the macroscopic observables, which we denote with a lower case notation $\hat o_j = \hat m_i^\alpha, \hat x$ or $\hat p$, we observe that $\overset{.}{K}(\hat o_j,\hat o_\ell) = f( h K(\hat o_p) K(\hat o_q,\hat o_r), h K(\hat o_p,\hat o_q, \hat o_r), hh'/N, h/(h'N))$ with $f$ a  linear function of its arguments, $p,q$ ranging from the possible observables of the system and $h,h'$ the possible coupling parameters ($\omega_{[...]}^{[...]}$, $g_\alpha$, $\kappa$ or $\Gamma$) (see SM for detailed calculations). The function has no independent first order cumulant terms. Therefore, assuming $h$ as finite coupling constants, given an initial uncorrelated  state (\textit{e.g.} a product state) with 
	\begin{equation}
		\label{eq.uncorrelatedMF.state}
		\lim_{N \rightarrow \infty}  K(\hat o_q,\hat o_r) = 0, \, \lim_{N \rightarrow \infty}  K(\hat o_p,\hat o_q,\hat o_r) = 0,
	\end{equation}
	 one has $\overset{.}{K}(\hat o_j,\hat o_\ell) = 0$ implying that the state remains uncorrelated, and therefore the dynamics shall be constrained to the MF first order cumulants, proving its exactness. We remark that despite MF is usually assumed exact for collective spin-boson systems, this may not always be the case. Recall that the derivative function $f$ also depends on the system couplings and its scaling with system size. 
	 Recently it was proposed collective spin systems acting as quantum heat engines for which the coupling strength scales non trivially with system size\cite{Leonardo2022}; despite still having a well defined thermodynamic limit, it leads to the failure of MF due to the unusual scaling and consequently to a nontrivial (enhanced) performance of the heat engine. 
	Our proof thus provides a simple understanding for these limitations, and can shed light for engineering more general systems failing MF with possibly unusual emergent features. 
	In summary, the MF approach is exact as long as the weight product between couplings and initial state is negligible according to the derivative function $f$ and Eq.\eqref{eq.uncorrelatedMF.state}. Moreover, since the resulting first order cumulants describe macroscopic observables, the corresponding dynamical rates in the Heisenberg equations of motion must be extensive with the system size (a subtler condition often not discussed~\cite{SM}).

	\textbf{\textit{Sufficient Condition.- }} For simplicity, we discuss here the case of a continuous time-independent Lindbladian. The case of Floquet Lindbladian follows similar reasoning, as we discuss later. The evolution of a quantum state $\hat \rho(t)$ with the Lindblad master equation is given by,
	\begin{equation}
		\hat \rho(t) = \hat \rho_{\rm NESS} + \sum_i e^{\lambda_i t} \mathcal{P}_i[\hat \rho(0)],
	\end{equation}
	where $\hat \rho_{\rm NESS}$ is the non-equilibrium steady state of the dynamics (i.e. $\mathcal{L}[\hat \rho_{\rm{NESS}}]=0$),  $\lambda_i$ are the generalized eigenvalues of the Lindbladian and $\mathcal{P}_i$ their corresponding superoperators. The gap of the Lindbladian is defined as,
	\begin{equation}
		\label{eq.gap}
		\Delta_N = \max_i \Re(\lambda_i),
	\end{equation}
	which are always nonpositive. Dissipative TC's breaking a continuous time-symmetry feature gapless excitations (i.e., $\lim_{N \rightarrow \infty} \Delta_N = 0$) along with a nonzero imaginary part for such eigenvalues ($\Im(\lambda_i)\neq 0 $) inducing nontrivial coherent oscillation in the system dynamics. 
	
	In the case of a non-degenerate Lindbladian ($\Delta_N \neq 0,\, \forall N$) we see directly that both limits commute $\lim_{N, t \rightarrow \infty} \hat \rho(t) = \lim_{t , N \rightarrow \infty} \hat \rho(t) = \hat \rho_{\rm NESS}$, where we use the notation $\lim_{A,B \rightarrow \infty }\equiv \lim_{A \rightarrow \infty }\lim_{B \rightarrow \infty }$. On the other hand, if the Lindbladian has gapless excitations in the thermodynamic limit - and only in this limit, thus excluding  possible decoherence-free subspaces  with $\Delta_N = 0$ for finite $N$ -  one may have a non-commutativity between these two limits,
	\begin{equation}
		\lim_{N, t \rightarrow \infty} \hat \rho(t) \neq \lim_{t , N \rightarrow \infty} \hat \rho(t).
	\end{equation}
	The non-commutativity of the NESS properties works as a sufficient condition for gapless modes in general non-degenerate Lindbladian, a main property we will explore in the manuscript. 
	
	In the thermodynamic limit the dynamics becomes exact within a mean-field approach, therefore a natural ``order-parameter'' to seek for the existence of gapless excitations 
	follows from the study of cumulant correlations. 
	Specifically,  given an initial uncorrelated state  $\lim_{t, N \rightarrow \infty}  K(\hat o_j, \hat o_\ell)(t) = 0$ (Eq.\eqref{eq.uncorrelatedMF.state}), a sufficient condition for non-commutativity follows from the inverse limit,
	\begin{equation}
		\label{eq.cum.cond.ness}
		\lim_{N , t \rightarrow \infty} K(\hat o_j, \hat o_\ell)(t) =
		\lim_{N \rightarrow \infty} [K(\hat o_j, \hat o_\ell)]_{\rm NESS} \neq 0,
	\end{equation}
	where $[K(\hat o_j, \hat o_\ell]_{\rm NESS} = 
	(\langle \hat o_j \hat o_\ell \rangle - \langle \hat o_i \rangle \langle \hat o_\ell \rangle)_{\rm NESS}$, 
	showing that nonzero macroscopic cumulant correlations in the NESS  come along with gapless excitations. 
	
	We focus here in the discussion of cumulants, but it is worth remarking that any other noncommuting feature could also be employed as a sufficient criterium for gapless modes. 
	For example, a class of states with null macroscopic cumulant correlations are those of coherent pure states - the mean-field pure state ansatz. Assuming that this relation is bijective, i.e., any state with null macroscopic cumulants corresponds to a coherent pure state, our gapless condition could be rephased in terms of any property not shared by coherent pure states, which may be simpler  to determine depending on the system. An equivalent condition could rely in this way on the purity of the state, as studied for spin systems with p-order-interactions~\cite{Giulia2021} or with a modified parity-time symmetry~\cite{Yuma2022}, for which the mixedness of the steady state was indeed observed in association to gapless modes and furthermore to the presence of boundary time-crystals.
	
	Considering Eq.\eqref{eq.cum.cond.ness} is satisfied the mean-field dynamics can behave in different forms due to the existence of conserved quantities in the system: (i) if the correlations concern to boson degrees of freedom, the mean-field may still reproduce the NESS one-body macroscopic observables correctly, i.e., $[K(\hat o_j)]_{\rm NESS} = [K(\hat o_j)]_{\rm MF}$ since there are no constraints to their expectation values; (ii) however, dealing with collective spins this can never be reached due to the angular momentum conservation $\sum_{\alpha} \langle (\hat m_i^\alpha)^2 \rangle = 1 $, and this is a crucial observation. Specifically, this conservation can be rewritten as 
	$\sum_{\alpha}[K(\hat m_i^\alpha)]^2_{\rm NESS}+[K(\hat m_i^\alpha,\hat m_i^\alpha)]_{\rm NESS} = 1 = \sum_{\alpha}[K(\hat m_i^\alpha)]^2_{\rm MF} $. Given Eq.\eqref{eq.cum.cond.ness} and from the fact that $K(\hat m_i^\alpha,m_i^\alpha)\geq 0$ $\forall j$, there must be at least an $\alpha'$ such that $[K(\hat m^{\alpha'}_i)]_{\rm NESS} \neq [K(\hat  m^{\alpha'}_i)]_{\rm MF}$ in the equality. Therefore, in this case the MF fails completely in the attempt to reproduce the one-body NESS macroscopic observables, and the  dynamics can  become ``lost'' due to its inability to match these expectation values, as illustrated in Fig.\eqref{fig:shell.su2}.  
	We study below different models for which the cumulant gapless condition is satisfied, and  find an interesting connection to persistent dynamics with the possible formation of dissipative time-crystal behavior~\cite{footnote1}. The inability of MF to reach the NESS values seems to lie at the core of these behaviors, showing the importance of conservation laws (as also considered in different models Refs.\cite{Angelo2017,Surace2019,Collado2021,Collado2022}) and correlations for such phases.
	
	\textbf{\textit{Hermitian Lindblad operators. - }} Perhaps the simplest case correspond to Lindbadians with Hermitian jump operators, for which the dissipation leads to a collective dephasing on the spins or boson degrees of freedom suppressing the off-diagonal terms in the density matrix with respect to their eigenstate basis. If the Lindbladian has no degeneracy for finite system sizes, given any initial state the dynamics is driven towards the maximally mixed state $\hat \rho_{\mathbb{I}} = \mathbb{I}/d$, with $d$ the normalization constant,  which by definition has nonzero macroscopic cumulant correlations. The steady state in this case is trivial for any strength of dissipation, with no specific ordering among the spins. This class of Lindbladians satisfy Eq.\eqref{eq.cum.cond.ness} and therefore always support gapless modes. A simple example follows a single spin ensemble ($M=1$) driven by a coherent field Hamiltonian along the $x$-direction ($\omega_x$) and a dissipation along the orthogonal $z$-direction ($\Gamma_{z,z}$), with all other parameters null in the Lindbladian (Eq.\eqref{eq.master.equation.spin}).  The gap eigenvalue $\lambda_1$ (the one with largest nonzero real part) of the Lindbladian shows in the limit of large system sizes a gapless scaling $-\Re(\lambda_1) \sim 1/N$ with an imaginary term $|\Im(\lambda_1)| \sim \omega_x$.
	While the decay rate (real part) arises from the dephasing, the coherent oscillations (imaginary term) follow directly from the field applied to the spins, with both features acting roughly independently of each other (see~\cite{SM}). Although one may still observe persistent dynamics on its observables, it arises simply from the applied field on the spins (notice that the frequency is independent of the dissipation), and not due to a correlated dynamics.
	A different situation arises, however, for systems with non-Hermitian Lindblad operators. In this case the competing coherent and dissipative dynamics can generate nontrivial steady states and possibly ordered dissipative time-crystal phases. We discuss examples below.

	\begin{figure}
		\includegraphics[width = 0.49 \linewidth]{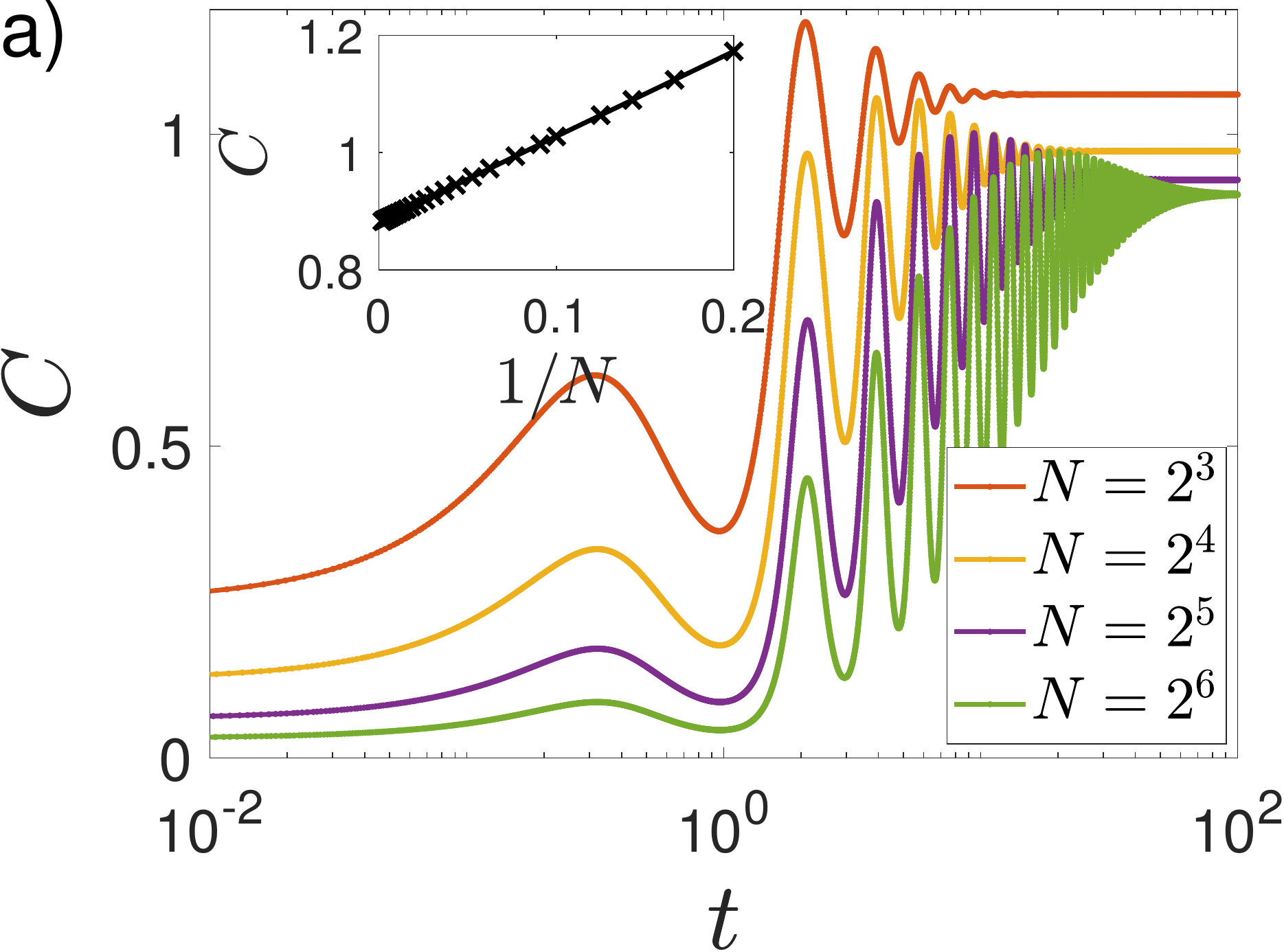}
		\includegraphics[width = 0.49 \linewidth]{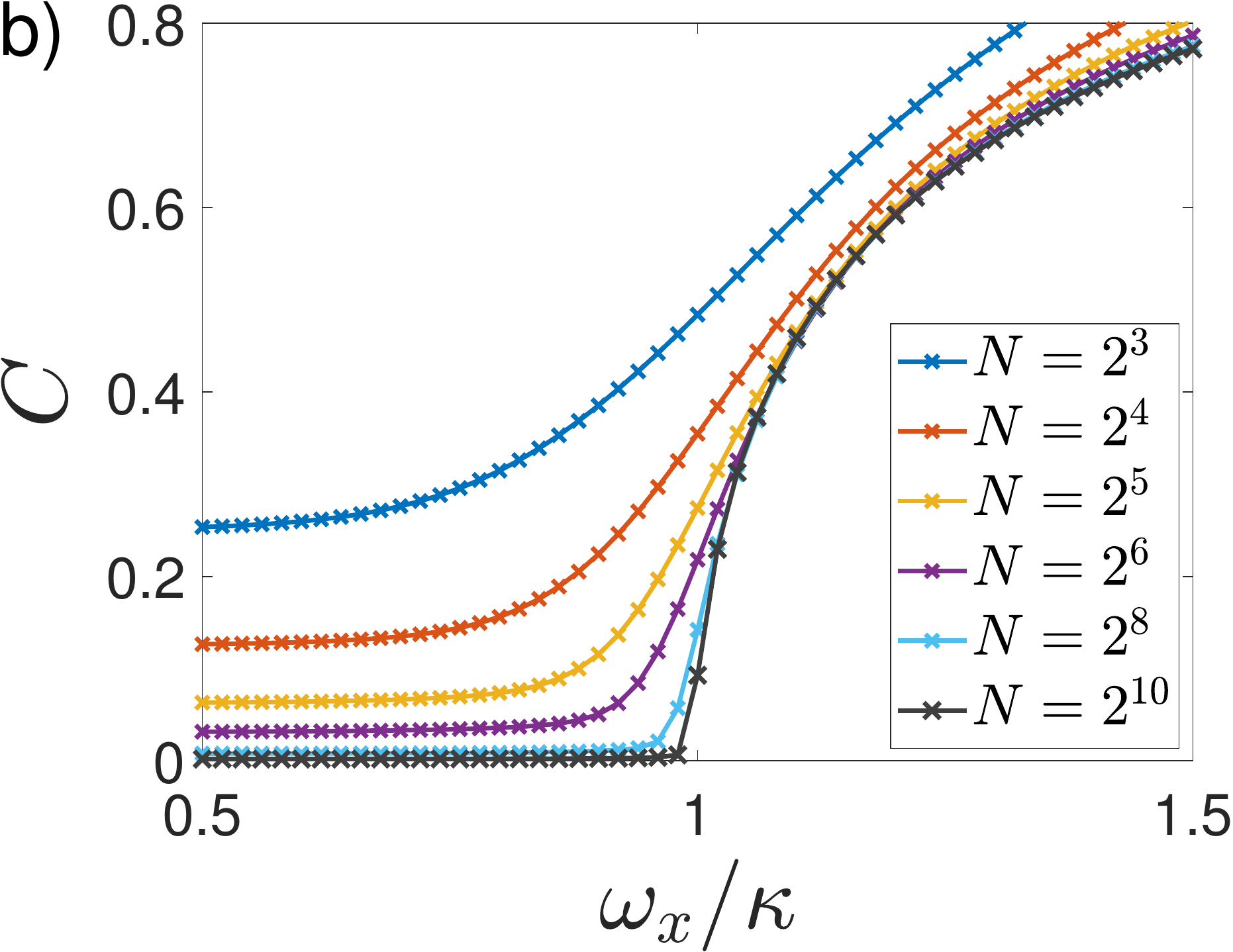}
		\caption{ \textbf{(a)} Dynamics of cumulants $C = \sum_\alpha K(\hat m_i^\alpha,\hat m_i^\alpha)$ for different system sizes, with $\omega_x/\kappa=2$. The inset shows the finite size scaling analysis, converging to a nonnull value in the thermodynamic limit.   \textbf{(b)} NESS cumulants for varying ratio $\omega_x/\kappa$ and system sizes.
		}
		\label{fig:cumulant.su2}
	\end{figure}
	
	\textbf{\textit{Collective spin-1/2 model. - }} A model with non-Hermitian Lindblad operators supporting nontrivial steady states  corresponds to a single spin ensemble ($M=1$) driven by competing coherent transverse field $\omega_x$ and a dissipative decay with \begin{equation}\label{eq:dissipation.su2}
		\Gamma_{x,x} = \Gamma_{y,y} = \sqrt{\kappa}, \quad  \Gamma_{y,x} = i \sqrt{\kappa} = \Gamma_{x,y}^*,
	\end{equation}
	(equivalently, the dissipation corresponds to a decay Lindblad jump operator   $\sqrt{\kappa} \hat S_-$, with $\hat S_\pm = (\hat S_x \pm i \hat S_y)$), while all other parameters are null. As discussed in Ref.\cite{Iemini2018} in the thermodynamic limit (and only in this limit) the model features persistent oscillations of its macroscopic magnetization (dissipative TC) for the weak dissipative regime $\omega_x/\kappa > 1 $ (see Fig.(\ref{fig:shell.su2})), while in the strong dissipative case $\omega_x/\kappa < 1 $ it shows a relaxation to its steady state. 
	We show in Fig.(\ref{fig:cumulant.su2}a) the dynamics for the cumulants in the dissipative TC phase with its oscillations and nonvanishing thermodynamic limit (inset). The gapless excitations tend to spread correlations among the spin constituents of the ensemble. In Fig.(\ref{fig:cumulant.su2}b) we compute the cumulant phase diagram for the steady state of the model. 
	While for stronger dissipation the spins dominantly decay, roughly pointing all down along the $z$-direction and therefore with null correlations among them, for weaker dissipation there are indeed nonzero macroscopic correlations in the NESS. The region with dissipative TC phase is therefore precisely the one with nonnull macroscopic cumulants, corroborating our arguments. We also analyzed extended models composed by a pair of interacting spin-$1/2$ systems ($M=2$). Our results show again a connection between cumulant correlations to dissipative TC's (see \cite{SM} for details).

\begin{figure}
	\includegraphics[width = 0.49 \linewidth]{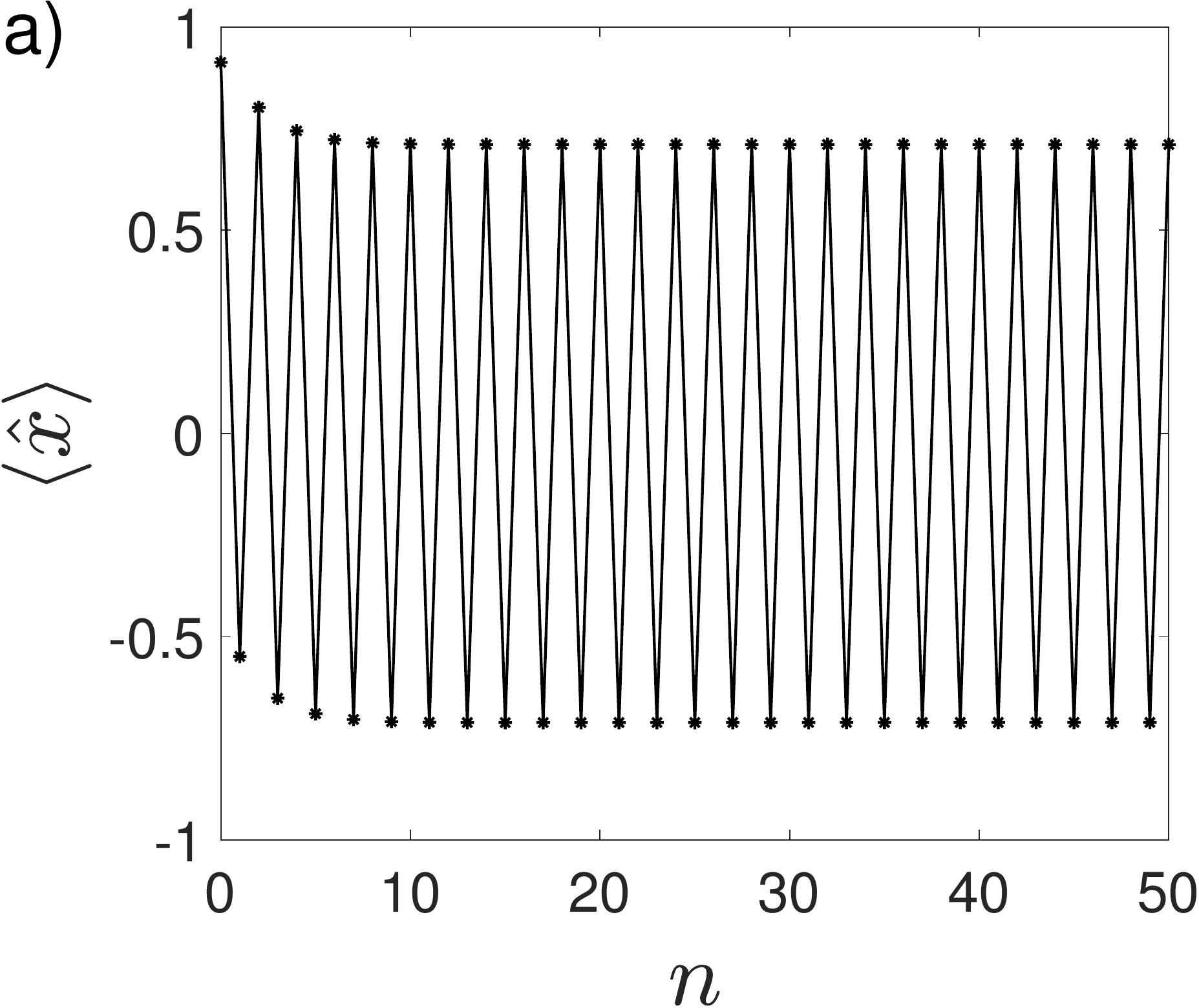}
	\includegraphics[width = 0.49 \linewidth]{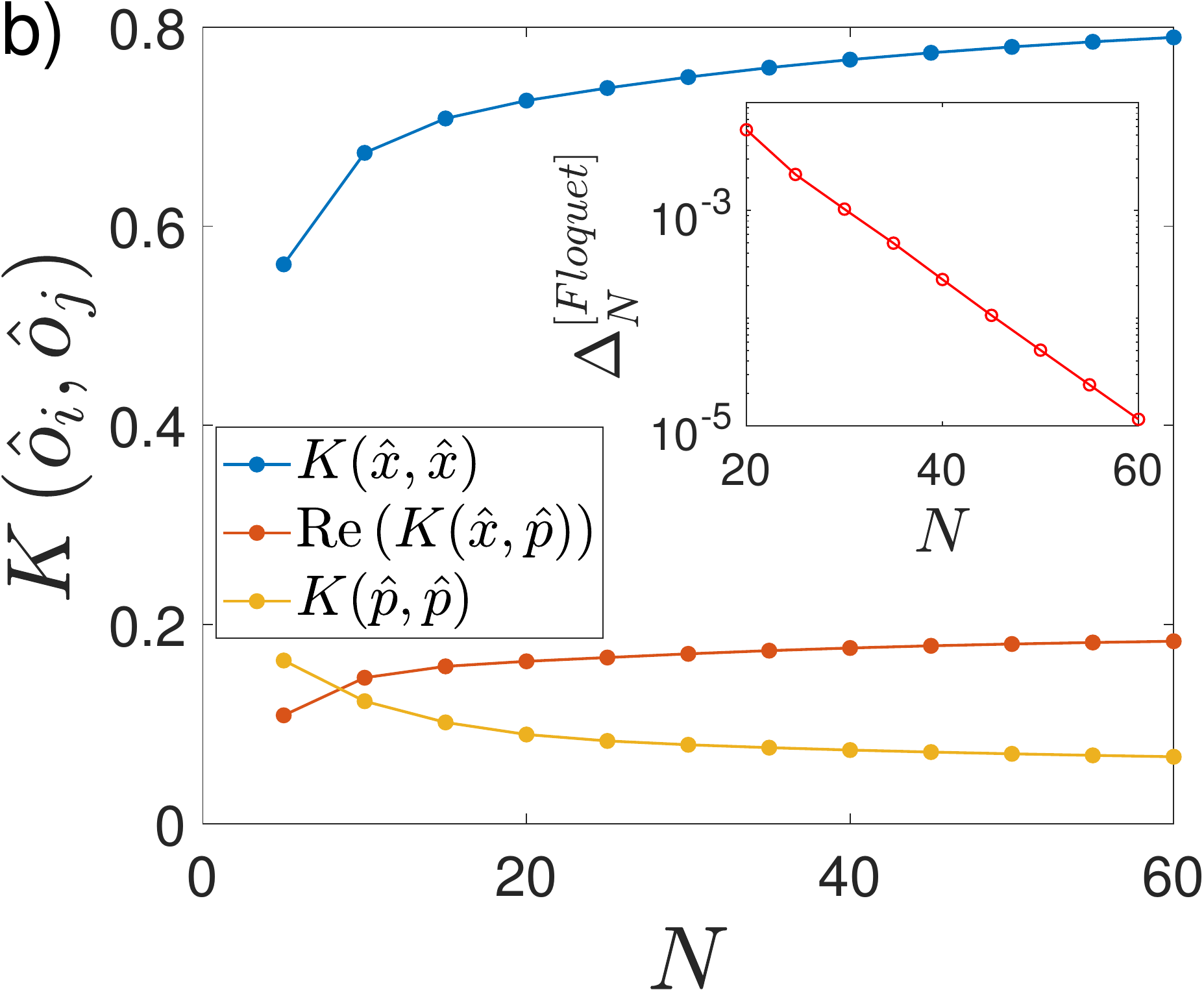}
	\caption{ {\bf (a)} Stroboscopic dynamics ($t_n=nT$)
		for the macroscopic position operator with $N=60$, considering  a  coherent state as initial condition. {\bf (b)} Cumulant correlations $K(\hat o_i,\hat o_j)$ in the NESS for different macroscopic observables. The inset shows the floquet spectral gap $\Delta_{N}^{[Floquet]}$ (Eq.\eqref{eq.floquet.gap}) with the number of spins $N$, highlighting its exponential decay. In all plots we used a maximum boson occupation of $d_b = 65$. 
	} 
	\label{fig:cumulant.Dicke.model}
\end{figure}

\textbf{\textit{Floquet spin-boson system.- }} We also consider time-dependent Floquet Lindbladian dynamics. Specifically, a modulated open Dicke model  with a single spin ensemble ($M=1$) interacting with a single-mode cavity. This model supports a dissipative time-crystals phase robust to perturbations, as examined in detail by Zongping Gong \textit{et al.} in ref.\cite{Gong2018}. The Lindbladian is defined as Eq.\eqref{eq.master.equation.spin} where the spin-boson coupling $g_x(t)$ is modulated periodically,
\begin{equation}
	g_x(t+T)=g_x(t)=
	\begin{cases}
		g, &  0\leq t \le \frac{T}{2}, \\
		0, & \frac{T}{2}\leq t \le T,
	\end{cases}
	\label{eq.modulated.lambda}
\end{equation}
with $T$ the Floquet period, the cavity (spin) field is given by $\omega_b$ ($\omega_z^{(1)}$) and cavity loss by the rate $\kappa_b$. All other parameters are null. The model can break its discrete time-translation symmetry showing subharmonic oscilations with period $nT$, for $n> 1$.

In the case of Floquet dynamics we must revisit the gap definition. It is now appropriate dealing with the
Floquet-Lindblad superoperator 
$\mathcal{U}_F = \mathcal{T} e^{\int_0^T dt \mathcal{L}(t)} $. The Floquet steady state is described by the eigenvalue of the operator $\mathcal{U}_F $ with $\lambda_i = 1$. The gap describing the slowest relaxation mode of the system is the closest to the unit ratio, i.e, 
\begin{equation} \label{eq.floquet.gap}
	\Delta_N^{[\rm Floquet]} =  1 - \max_{\{ \lambda_i\}}|\lambda_i|.
\end{equation}
With this definition all our gapless condition arguments  remain unchanged, based on the cumulant correlations of the Floquet steady state. 

The analysis of the dynamics for the system is very demanding in the limit of large number of spins due to the high dimensionality of the Hilbert space. Nevertheless, one can work with an effective description of the model, within an adiabatic elimination of the spin degrees of freedom and under specific conditions~\cite{Gong2018}. The boson-only description is given by the coherent Hamiltonian,
\begin{align}
	&\hat H = \omega_b \hat a^{\dagger}\hat a-\frac{\Omega_2(t)}{4}\left(\hat a^{\dagger}+\hat a\right)^2+\frac{\Omega_4(t)}{32N}\left(\hat a^{\dagger}+\hat a\right)^4,
	\label{eq.bosonic.effetive.dynamics}
\end{align}
and losses as Eq.\eqref{eq.dissipation.boson}. The couplings are given by,
\begin{equation}
	\Omega_2(t+T)=\Omega_2(t)=
	\begin{cases}
		1.5\omega_b , &  0\leq t \le \frac{\pi}{\omega_b }, \\
		0, & \frac{\pi}{\omega_b }\leq t \le \left(2-\epsilon\right)\frac{\pi}{\omega_b },
	\end{cases}
	\label{eq.modulated.Omega}
\end{equation}
with $\Omega_{4}(t)=\Omega_{2}(t)$. The dynamics of the system can be studied numerically, truncating the maximum boson occupancy sufficiently large. Given an initial coherent state, we show in Fig.(\ref{fig:cumulant.Dicke.model}a) the Floquet dynamics for the system in the thermodynamic limit, with couplings $\kappa_b=0.1\omega_b $ and $\epsilon=0.1$ supporting a dissipative TC. 
The macroscopic position operator, after an initial transient time, is characterized by a stable period doubling dynamics. 
Performing a finite-size scaling for the macroscopic cumulant correlations we obtain their scaling with system size shown in Fig.(\ref{fig:cumulant.Dicke.model}b). Once more, concomitant with macroscopic correlations the system supports gapless Floquet excitations, as shown in the inset of Fig.(\ref{fig:cumulant.Dicke.model}b).

\textbf{\textit{Permutationally invariant systems.-}} Our MF proof and gapless condition can be extended to more general systems, as we illustrate below. Specifically, we generalize to Lindbladians satisfying a weaker restriction, that the expectation values of the spins be permutationally invariant in the dynamics:
\begin{equation}
	\biggl< \prod_{j=1}^n \hat \sigma_{i,k_j}^{\alpha_j} \biggr> = \biggl< \prod_{j=1}^n \hat \sigma_{i,k'_j}^{\alpha_j} \biggr>, \quad \forall k_j, k'_j, i, n,
\end{equation}
where $\langle \hat O \rangle \equiv Tr(\hat \rho(t) \hat O)$, $k_p\neq k_{q}$ and $k_{p}^{\prime}\neq k_{q}^{\prime}$, for ${p}\neq{q}$, i.e., for each $i$'th spin ensemble, the $n$-body correlated magnetization is independent on the spin microscopic labels ($k_j$ and $k_{j}'$). We consider as an example a dissipative spin channel with possible spatial dependence,
\begin{equation}\label{eq.master.equation.spin.perm.inv}
	\mathcal{D}_i[\hat \rho] = \sum_{j,k=1}^N\sum_{\alpha,\beta}\gamma_{\alpha,\beta}^{(i)\, j,k} \left(\hat \sigma_{i,j}^{\alpha} \hat\rho  \hat \sigma_{i,k}^{\beta} - \frac{1}{2}\{\hat \sigma_{i,k}^{\beta} \hat \sigma_{i,j}^{\alpha},\hat\rho \}\right),
\end{equation}
with well defined thermodynamic couplings, i.e., 
$\lim_{N \rightarrow \infty} \sum_{j=1}^N\gamma_{ \alpha,\beta}^{(i)\, j,j}/N$ and $\lim_{N \rightarrow \infty}  \sum_{j\neq k}^N \gamma_{\alpha,\beta}^{(i)\,j,k}/N$ finite.
The previously discussed collective spin dissipation (Eq.\eqref{eq.dissipation.spin}), as well sufficiently long-range~\cite{Gianluca2022} and strictly local ones ($\gamma_{\alpha, \beta}^{i,j} = \delta_{i,j} \tilde{\gamma}_{\alpha, \beta}$) lie as specific cases of the above channel~\cite{Nori2018}. The proof follows similarly to the previous discussions (see~\cite{SM} for details).

\textbf{\textit{Conclusion.- }} In this work we discussed a condition for gapless excitations in a class of Lindbladians described by collective spin-boson models and general permutationally invariant systems. The condition based only on the macroscopic cumulant correlations is important not only for analytically establishing the fundamental relationship between NESS properties and system dynamics, but also from a practical point of view in the spectral determination without the need for a time evolution integration  of the master equation neither its exact diagonalization.   These results can be insightful towards a proper understanding of the basic mechanisms for persistent dynamics in the thermodynamic limit. 
Interesting perspectives rely on the generalization for different models, e.g. nonpermutationally invariant systems~\cite{joseph2022}. Moreover, since our condition is a sufficient one, it would be important to extend to the necessary conditions for gapless excitations as well. Finally, we expect that our results will have an impact on the experimental implementation. Such phases have recently been observed in atom-cavity systems~\cite{hans2019,hans2020,hans2021,Phatthamon2022}. Envisioned with state-of-the-art quantum simulation platforms~\cite{lorenza2001,Fink2009,Minghui2014,Shankar2017,Gangloff2019}
 one could engineer different forms of correlated dissipative steady states in spin-boson systems, and therefore different dissipative time-crystal implementations.

\textbf{\textit{Acknowledgments.- }} We acknowledge enlightening discussions with A. Russomanno, G. Passarelli, F. Carollo and I. Lesanovsky. The authors acknowledge financial support from the Brazilian funding agencies CAPES, CNPQ and FAPERJ (Grant No. 308205/2019-7, No. E-26/211.318/2019, No. 151064/2022-9 and E-26/201.365/2022). The codes for the numerical simulations have been constructed using open source QuTiP library and free software Octave\cite{footnote2}.

\widetext
\clearpage

\begin{center}
	\large \textbf{Supplemental Material} \\ \vspace{0.3cm} 
	Leonardo da Silva Souza, Luis Fernando dos Prazeres and Fernando Iemini
\end{center}
\setcounter{section}{0}
\setcounter{equation}{0}
\setcounter{figure}{0}
\setcounter{table}{0}
\setcounter{page}{1}
\makeatletter
\renewcommand{\theequation}{S\arabic{equation}}
\renewcommand{\thefigure}{S\arabic{figure}}
\renewcommand{\bibnumfmt}[1]{[S#1]}
\renewcommand{\citenumfont}[1]{S#1}

In this Supplemental Material we give more details on the derivation of the MF semiclassical equations, the proof for its exactness within cumulant approach and the study of Hermitian Lindblad operators.

\section*{Cumulant expansion}

In this section we discuss the cumulant expansion, a concept originally develop for the characterization of classical stochastic variables \cite{Gardiner1985} which has been generalized to operators by Kubo \cite{Kubo1962}. The cumulant of $n$ operators $\{O_1,\dots,O_n\}$ can be written as 
\begin{equation}
	K\left(O_{1},...,O_{N}\right)=\sum_{\pi\in P}\left(\left|\pi\right|-1\right)!(-1)^{\left|\pi\right|-1}\prod_{B\in\pi}\langle\prod_{j\in B}O_{j}\rangle,
	\label{cumulant.expansion}
\end{equation}
where $P$ is the set of all partition in $\{1,2,\dots,n\}$, $|\pi|$ is the length of the partition $\pi$ and $B$ runs over the blocks of partition $\pi$.
Cumulants share important properties that make them  interesting quantifiers for correlations: (i) the cumulant of  a set of operators is null if any one or more operators in the set are statistically independent to each other~\cite{Kubo1962}; (ii) any $n$'th order correlation ($\langle \hat O_1 ... \hat O_n\rangle$) can be decomposed in terms of products and sums of lower orders $n'<n$ correlations, plus the corresponding $n$'th order cumulant. Therefore, the cumulant can be seem as the irreducible part of the $n$'th order correlation (the term that cannot be decomposed into lower ones). As an illustration, in the case of second and third order correlations we have,
\begin{eqnarray}
	\langle \hat O_j \hat O_\ell \rangle &=& \langle\hat O_j\rangle\langle\hat O_\ell\rangle+K(\hat O_j,\hat O_\ell), \nonumber \\
	\langle \hat O_j \hat O_\ell \hat O_m\rangle &=& \langle \hat O_j \hat O_\ell\rangle\langle\hat O_m\rangle+\langle \hat O_j \hat O_m\rangle\langle\hat O_\ell\rangle +\langle\hat O_\ell \hat O_m\rangle\langle\hat O_j\rangle-2\langle \hat O_j\rangle\langle \hat O_\ell\rangle\langle \hat O_m\rangle+K(\hat O_j,\hat O_\ell, \hat O_m).
\end{eqnarray}
An  interesting application of cumulants approach is that it allows us to derive effective closed dynamical equations of motion for the system observables, as follows. For systems with up to two body Hamiltonian correlations, and one-body jump operators, the Heisenberg equations of motion for $n$'th order correlations depends on terms up to $n+1$ order, thus leading to an infinite hierarchical set of equations of motion in order to fully characterize the dynamics. Nevertheless, once we truncate the cumulant correlations at a $k$'th order (i.e., $K(\hat O_{i_1},...,\hat O_{i_k})=0,\, \forall i$) and obtain an effective closed set of equations of motion for the correlations up to $k-1$'th order. E.g, the  MF approximation corresponds a truncation at the second order cumulant, i.e., discarding all correlations between two operators. Vanishing of $k>2$ order cumulants lead to a Gaussian distribution for the observables~\cite{Wunsche2015}. Therefore, cumulant expansion allows an approach to approximate the solution for the dynamics. In many scenarios it can even provide an exact solution as it will be discussed later.

\section*{Semiclassical Limit} 
\label{d2.dynamical.equations}

We derive the dynamical equations of motion for its observables and employ a semiclassical approach, usual to collective systems.
The equations of motion of any operator $\hat o$ within the Heisenberg picture is given by, 
\begin{equation}
	\frac{ d \langle \hat o \rangle }{dt} = i \langle [\hat H,\hat o] \rangle + \frac{1}{2S}\sum_{i=1}^M\sum_{\alpha,\beta}\Gamma_{\alpha,\beta}^{(i)} 
	\langle [\hat S_{i}^{\beta},\hat o ] \hat S_{i}^{\alpha}
	+ \hat S_{i}^{\beta}[\hat o ,\hat S_{i}^{\alpha}] \rangle+ \frac{\kappa}{2} 
	\langle [\hat{a}^{\dagger},\hat o ] \hat{a} + \hat{a}^{\dagger}[ \hat o ,\hat{a}] \rangle,
	\label{eq:dynamical.equation.chain}
\end{equation}
where $\langle\cdot\rho\rangle=\text{Tr}\left[\cdot\rho\right]$. Taking for $\hat o$ the the macroscopic observables $\hat x =\frac{\hat a + \hat{a}^{\dagger}}{\sqrt{2N\omega_b}}$, $\hat p =i\frac{\hat{a}^{\dagger}-\hat a}{\sqrt{2N/\omega_b}}$, $\hat m_{j}^\alpha =\hat S_{j}^\alpha/S$ and using SU(2) commutation relations, $[\hat S_{i}^\alpha,\hat S_{j}^\beta] = i\epsilon^{\alpha \beta \gamma}\hat S_{i}^\gamma\delta_{i,j}$, and bosonic algebra, $\left[\hat{a},\hat{a}^{\dagger}\right]=1$. We can define the corresponding dynamical equation,
\begin{align}
	\frac{d}{dt} \langle \hat{x} \rangle 
	&=\langle\hat{p}\rangle-\frac{\kappa}{2}\langle\hat{x}\rangle,\nonumber\\
	\frac{d}{dt} \langle \hat{p} \rangle 
	&=-\omega_{b}^2\langle\hat{x}\rangle-\frac{\kappa}{2}\langle\hat{p}\rangle-\sum_{i=1}^{M}\sum_{\alpha}\frac{g_{\beta}\sqrt{2\omega_{b}}}{2}\langle\hat{m}_{i}^{\alpha}\rangle,\nonumber\\
	\frac{d}{dt} \langle \hat m_{j}^\gamma \rangle 
	&=-\sum_{\alpha}\epsilon^{\alpha\gamma\tau}\left(\omega_{\alpha}^{(j)}\langle\hat m_{j}^{\tau}\rangle+g_{\alpha}\sqrt{2\omega_{b}}\langle\hat{x}\hat{m}_{j}^{\tau}\rangle\right)\\
	&\quad-\sum_{i=1}^{M}\sum_{\alpha,\beta} \left(\epsilon^{\beta\gamma\tau}\omega_{\alpha,\beta}^{(i,j)}\langle \hat m_{i}^{\alpha} \hat m_{j}^{\tau}\rangle+\epsilon^{\alpha\gamma\tau}\omega_{\alpha,\beta}^{(j,i)}\langle \hat m_{j}^{\tau}\hat m_{i}^{\beta}\rangle\right)\nonumber\\ &\quad+\frac{1}{2}\sum_{\alpha,\beta}i\Gamma_{\alpha,\beta}^{(j)} 
	\left( \epsilon^{\beta\gamma\tau}\langle\hat m_{j}^{\tau}\hat m_{j}^{\alpha}\rangle
	+ \epsilon^{\gamma\alpha\tau}\langle\hat m_{j}^{\beta}\hat m_{j}^{\tau}\rangle\right).\nonumber
\end{align}

In the mean-field approach, we assume mean values factorable (e.g. $\langle \hat x \hat m^\alpha \rangle \cong \langle \hat x \rangle \langle \hat m^\alpha \rangle$). The equations of motion can be written as
\begin{align}
	\frac{d}{dt}x 
	&=p-\frac{\kappa}{2}x,\nonumber\\
	\frac{d}{dt}p 
	&=-\omega_{b}^2x-\frac{\kappa}{2}p-\sum_{i=1}^{M}\sum_{\alpha}\frac{g_{\beta}\sqrt{2\omega_{b}}}{2}m_{i}^{\alpha},\nonumber\\
	\frac{d}{dt} m_{j}^\gamma  
	&=-\sum_{\alpha}\epsilon^{\alpha\gamma\tau}\left(\omega_{\alpha}^{(j)} m_{j}^{\tau}+g_{\alpha}\sqrt{2\omega_{b}}xm_{j}^{\tau}\right)\\
	&\quad-\sum_{i=1}^{M}\sum_{\alpha,\beta} \left(\epsilon^{\beta\gamma\tau}\omega_{\alpha,\beta}^{(i,j)}m_{i}^{\alpha} m_{j}^{\tau}+\epsilon^{\alpha\gamma\tau}\omega_{\alpha,\beta}^{(j,i)}m_{j}^{\tau} m_{i}^{\beta}\right)\nonumber\\ &\quad+\frac{1}{2}\sum_{\alpha,\beta}i\Gamma_{\alpha,\beta}^{(j)} 
	\left( \epsilon^{\beta\gamma\tau}m_{j}^{\tau}m_{j}^{\alpha}
	+ \epsilon^{\gamma\alpha\tau} m_{j}^{\beta} m_{j}^{\tau}\right),\nonumber
\end{align}
where we use $m_{j}^\alpha \equiv \langle \hat m_{j}^\alpha \rangle$ to simplify our notation. These equations conserves the norm $(m_{j}^x)^2 +(m_{j}^y)^2+ ( m_{j}^z)^2=1$, with $j=1,\dots,n$.

\section*{Exactness of semiclassical dynamics}

In this section we discuss in detail the exactness of the mean-field approach in the thermodynamic limit for the model. We show that for  certain conditions, such as initial states closed in the second cumulants, the temporal derivatives of the second cumulants are null.
First, in order to compute the dynamical equations of the second cumulants for the observables of interest, we define the two-body equations of motion

\begin{align}
	\frac{d}{dt} \langle \hat{x}^2\rangle 
	&=\langle\hat{x}\hat{p}+\hat{p}\hat{x}\rangle-\kappa\langle\hat{x}^2\rangle +\frac{\kappa}{2N\omega_{b}},\nonumber\\
	\frac{d}{dt} \langle \hat{x}\hat{p} \rangle 
	&=-\omega_{b}^2\langle\hat{x}^2\rangle+\langle\hat{p}^2\rangle-\frac{\kappa}{2}\langle\hat{x}\hat{p}+\hat{p}\hat{x}\rangle-\sum_{i=1}^{M}\sum_{\alpha}\frac{g_{\alpha}\sqrt{2\omega_{b}}}{2}\langle\hat{x}\hat{m}_{i}^{\alpha}\rangle,\nonumber\\
	\frac{d}{dt} \langle \hat{p}^2 \rangle 
	&=-\omega_{b}^2\langle\hat{x}\hat{p}+\hat{p}\hat{x}\rangle-\kappa\langle\hat{p}^2\rangle-\sum_{i=1}^{M}\sum_{\alpha}g_{\alpha}\sqrt{2\omega_b}\langle\hat{p}\hat{m}_{i}^{\alpha}\rangle+\frac{\kappa\omega_{b}}{2N},\nonumber
\end{align}
\begin{align}
	\frac{d}{dt} \langle \hat{x}\hat{m}_{j}^{\gamma} \rangle
	&=\langle\hat{p}\hat{m}_{j}^{\gamma}\rangle-\frac{\kappa}{2}\langle\hat x\hat{m}_{j}^{\gamma}\rangle-\sum_{\alpha}\epsilon^{\alpha\gamma\tau}g_{\alpha}\sqrt{2\omega_{b}}\langle\hat{x}^2\hat{m}_{j}^{\tau}\rangle\nonumber\\
	&\quad-\sum_{\alpha}\epsilon^{\alpha\gamma\tau}\omega_{\alpha}^{(j)}\langle\hat{x}\hat m_{j}^{\tau}\rangle-\sum_{i=1}^{M}\sum_{\alpha,\beta} \left(\epsilon^{\beta\gamma\tau}\omega_{\alpha,\beta}^{(i,j)}\langle \hat x\hat m_{i}^{\alpha} \hat m_{j}^{\tau}\rangle+\epsilon^{\alpha\gamma\tau}\omega_{\alpha,\beta}^{(j,i)}\langle\hat x \hat m_{j}^{\tau}\hat m_{i}^{\beta}\rangle\right)\nonumber\\
	&\quad+\frac{1}{2}\sum_{\alpha,\beta}i\Gamma_{\alpha,\beta}^{(j)} 
	\left( \epsilon^{\beta\gamma\tau}\langle\hat x\hat m_{j}^{\tau}\hat m_{j}^{\alpha}\rangle
	+ \epsilon^{\gamma\alpha\tau}\langle\hat x\hat m_{j}^{\beta}\hat m_{j}^{\tau}\rangle\right),
\end{align}
\begin{align}
	\frac{d}{dt} \langle \hat{p}\hat{m}_{j}^{\gamma} \rangle
	&=-\omega^2\langle\hat{x}\hat{m}_{j}^{\gamma}\rangle-\frac{\kappa}{2}\langle\hat{p}\hat{m}_{j}^{\gamma}\rangle-\sum_{\alpha}\epsilon^{\alpha\gamma\tau}g_{\alpha}\sqrt{2\omega_{b}}\langle\hat{x}\hat{p}\hat{m}_{j}^{\tau}\rangle-\sum_{i=1}^M\sum_{\alpha}\frac{g_{\alpha}\sqrt{2\omega_b}}{2}\langle\hat{m}_{i}^{\alpha}\hat{m}_{j}^{\gamma}\rangle\nonumber\\
	&\quad-\sum_{\alpha}\epsilon^{\alpha\gamma\tau}\omega_{\alpha}^{(j)}\langle\hat{p}\hat m_{j}^{\tau}\rangle-\sum_{i=1}^{M}\sum_{\alpha,\beta} \left(\epsilon^{\beta\gamma\tau}\omega_{\alpha,\beta}^{(i,j)}\langle \hat p\hat m_{i}^{\alpha} \hat m_{j}^{\tau}\rangle+\epsilon^{\alpha\gamma\tau}\omega_{\alpha,\beta}^{(j,i)}\langle\hat p \hat m_{j}^{\tau}\hat m_{i}^{\beta}\rangle\right)\nonumber\\
	&\quad+\frac{1}{2}\sum_{\alpha,\beta}i\Gamma_{\alpha,\beta}^{(j)} 
	\left( \epsilon^{\beta\gamma\tau}\langle\hat p\hat m_{j}^{\tau}\hat m_{j}^{\alpha}\rangle
	+ \epsilon^{\gamma\alpha\tau}\langle\hat p\hat m_{j}^{\beta}\hat m_{j}^{\tau}\rangle\right),\nonumber
\end{align}
\begin{align}
	\frac{d}{dt} \langle\hat m_{j}^\gamma\hat m_{\ell}^\zeta\rangle 
	&=-\sum_{\alpha}\left(\epsilon^{\alpha\gamma\tau}\omega_{\alpha}^{(j)}\langle\hat m_{j}^{\tau}\hat m_{\ell}^\zeta\rangle+\epsilon^{\alpha\zeta\tau} \omega_{\alpha}^{(\ell)}\langle\hat m_{j}^{\gamma} \hat m_{\ell}^{\tau}\rangle\right)-\sum_{\alpha}g_{\alpha}\sqrt{2\omega_{b}}\left(\epsilon^{\alpha\gamma\tau}\langle\hat x\hat m_{j}^{\tau}\hat m_{\ell}^\zeta\rangle+\epsilon^{\alpha\zeta\tau} \langle\hat x\hat m_{j}^{\gamma} \hat m_{\ell}^{\tau}\rangle\right)\nonumber\\
	&\quad-\sum_{i=1}^M\sum_{\alpha,\beta}\left(\epsilon^{\beta\gamma\tau}\omega_{\alpha\beta}^{(i,j)}\langle\hat m_{i}^{\alpha}\hat m_{j}^{\tau}m_{\ell}^{\zeta}\rangle+\epsilon^{\alpha\gamma\tau}\omega_{\alpha\beta}^{(j,i)}\langle\hat m_{j}^{\tau}\hat m_{i}^{\beta}\hat m_{\ell}^{\zeta}\rangle+\epsilon^{\beta\zeta\tau}\omega_{\alpha\beta}^{(i,\ell)}\langle\hat m_{j}^{\gamma}\hat m_{i}^{\alpha}m_{\ell}^{\tau}\rangle+\epsilon^{\alpha\zeta\tau}\omega_{\alpha\beta}^{(\ell,i)}\langle\hat m_{j}^{\gamma}\hat m_{\ell}^{\tau}m_{i}^{\beta}\rangle\right)\nonumber\\
	&\quad+\frac{i}{2}\sum_{\alpha,\beta}\left(\Gamma_{\alpha,\beta}^{(j)}\left(\epsilon^{\beta\gamma\tau}\langle\hat m_{j}^{\tau}\hat m_{\ell}^{\zeta}\hat m_{j}^{\alpha}\rangle+\epsilon^{\gamma\alpha\tau}\langle\hat m_{j}^{\beta}\hat m_{j}^{\tau}\hat m_{\ell}^{\zeta}\rangle\right)+\Gamma_{\alpha,\beta}^{(\ell)}\left(\epsilon^{\beta\zeta\tau}\langle\hat m_{j}^{\gamma}\hat m_{\ell}^{\tau}\hat m_{\ell}^{\alpha}\rangle+\epsilon^{\zeta\alpha\tau}\langle\hat m_{\ell}^{\beta}\hat m_{j}^{\gamma}\hat m_{\ell}^{\tau}\rangle\right)\right).\nonumber
\end{align}

These equations are derived using Eq.(\ref{eq:dynamical.equation.chain}). We can write the dynamical equations for $2^{nd}$ order cumulant,  $K\left(\hat o_{j},\hat o_{k}\right)=\langle\hat o_{j}\hat o_{k}\rangle-\langle\hat o_{j}\rangle\langle\hat o_{k}\rangle$, as follows
\begin{align}
	\frac{d}{dt}  K\left(\hat{x},\hat{x}\right) 
	&=K\left(\hat{x},\hat{p}\right)+K\left(\hat{p},\hat{x}\right)-\kappa K\left(\hat{x},\hat{x}\right)+ \frac{\kappa}{2N\omega_b},\nonumber\\
	\frac{d}{dt} K\left(\hat{x},\hat{p}\right) 
	&=-\omega_{b}^{2}K\left(\hat{x},\hat{x}\right)+K\left(\hat{p},\hat{p}\right)-\frac{\kappa}{2} \left(K\left(\hat{x},\hat{p}\right)+K\left(\hat{p},\hat{x}\right)\right)-\sum_{i=1}^M\sum_{\alpha}\frac{g_{\alpha}\sqrt{2\omega_b}}{2}K\left(\hat{x},\hat{m}_{i}^{\alpha}\right),\nonumber\\
	\frac{d}{dt} K\left(\hat{p},\hat{p}\right) 
	&=-\omega_{b}^{2}\left(K\left(\hat{x},\hat{p}\right)+K\left(\hat{p},\hat{x}\right)\right)-\kappa K\left(\hat{p},\hat{p}\right) -\sum_{i=1}^M\sum_{\alpha}g_{\alpha}\sqrt{2\omega_b}K\left(\hat{p},\hat{m}_{i}^{\alpha}\right)+ \frac{\kappa\omega_b}{2N},\nonumber
\end{align}
\begin{align}
	\frac{d}{dt} K\left(\hat{x},\hat{m}_{j}^{\gamma}\right) 
	&=K\left(\hat{p},\hat{m}_{j}^{\gamma}\right)-\frac{\kappa}{2}K\left(\hat{x},\hat{m}_{j}^{\gamma}\right)-\sum_{\alpha}\epsilon^{\alpha\gamma\tau}\omega_{\alpha}^{(j)}K\left(\hat p\hat m_{j}^{\tau}\right)\nonumber\\
	&\quad-\sum_{i=1}^{M}\sum_{\alpha,\beta} \left(\epsilon^{\beta\gamma\tau}\omega_{\alpha,\beta}^{(i,j)}\left(K\left(\hat x,\hat m_{i}^{\alpha}, \hat m_{j}^{\tau}\right)+K\left(\hat x,\hat m_{i}^{\alpha}\right)K\left(\hat m_{j}^{\tau}\right)+K\left(\hat x,\hat m_{j}^{\tau}\right)K\left(\hat m_{i}^{\alpha}\right)\right)\right.\nonumber\\
	&\qquad\qquad\quad+\left.\epsilon^{\alpha\gamma\tau}\omega_{\alpha,\beta}^{(j,i)}\left(K\left(\hat x, \hat m_{j}^{\tau},\hat m_{i}^{\beta}\right)+K\left(\hat x, \hat m_{j}^{\tau}\right)K\left(\hat m_{i}^{\beta}\right)+K\left(\hat x, \hat m_{i}^{\beta}\right)K\left(\hat m_{j}^{\tau}\right)\right)\right)\nonumber\\
	&\quad-\sum_{\alpha}\epsilon^{\alpha\gamma\tau}g_{\alpha}\sqrt{2\omega_b}\left(K\left(\hat x,\hat x,\hat{m}^{\tau}\right)+K\left(\hat x,\hat x\right)K\left(\hat{m}^{\tau}\right)+K\left(\hat x,\hat{m}^{\tau}\right)K\left(\hat x\right)\right)\nonumber\\
	&\quad+\frac{1}{2}\sum_{\alpha,\beta}i\Gamma_{\alpha,\beta}^{(j)} 
	\left( \epsilon^{\beta\gamma\tau}\left(K\left(\hat x,\hat m_{j}^{\tau},\hat m_{j}^{\alpha}\right)+K\left(\hat x,\hat m_{j}^{\tau}\right)K\left(\hat m_{j}^{\alpha}\right)+K\left(\hat x,\hat m_{j}^{\alpha}\right)K\left(\hat m_{j}^{\tau}\right)\right)\right.\nonumber\\
	&\qquad\qquad\qquad\left.+\epsilon^{\gamma\alpha\tau}\left(K\left(\hat x,\hat m_{j}^{\beta},\hat m_{j}^{\tau}\right)+K\left(\hat x,\hat m_{j}^{\beta}\right)K\left(\hat m_{j}^{\tau}\right)+K\left(\hat x,\hat m_{j}^{\tau}\right)K\left(\hat m_{j}^{\beta}\right)\right)\right),\nonumber
\end{align}
\begin{align}
	\frac{d}{dt} K\left(\hat{p},\hat{m}_{j}^{\gamma}\right) 
	&=-\omega_{b}^{2}K\left(\hat{x},\hat{m}_{j}^{\gamma}\right)-\frac{\kappa}{2}K\left(\hat{p},\hat{m}_{j}^{\gamma}\right)-\sum_{\alpha}\epsilon^{\alpha\gamma\tau}\omega_{\alpha}^{(j)}K\left(\hat p\hat m_{j}^{\tau}\right)-\sum_{i=1}^M\sum_{\alpha}\frac{g_{\alpha}\sqrt{2\omega_b}}{2}K\left(\hat{m}_{i}^{\alpha},\hat{m}_{j}^{\gamma}\right)\nonumber\\
	&\quad-\sum_{i=1}^{M}\sum_{\alpha,\beta} \left(\epsilon^{\beta\gamma\tau}\omega_{\alpha,\beta}^{(i,j)}\left(K\left(\hat p,\hat m_{i}^{\alpha}, \hat m_{j}^{\tau}\right)+K\left(\hat p,\hat m_{i}^{\alpha}\right)K\left(\hat m_{j}^{\tau}\right)+K\left(\hat p,\hat m_{j}^{\tau}\right)K\left(\hat m_{i}^{\alpha}\right)\right)\right.\nonumber\\
	&\qquad\qquad\quad+\left.\epsilon^{\alpha\gamma\tau}\omega_{\alpha,\beta}^{(j,i)}\left(K\left(\hat p, \hat m_{j}^{\tau},\hat m_{i}^{\beta}\right)+K\left(\hat p, \hat m_{j}^{\tau}\right)K\left(\hat m_{i}^{\beta}\right)+K\left(\hat p, \hat m_{i}^{\beta}\right)K\left(\hat m_{j}^{\tau}\right)\right)\right)\nonumber\\
	&\quad-\sum_{\alpha}\epsilon^{\alpha\gamma\tau}g_{\alpha}\sqrt{2\omega_b}\left(K\left(\hat x,\hat p,\hat{m}^{\tau}\right)+K\left(\hat x,\hat p\right)K\left(\hat{m}^{\tau}\right)+K\left(\hat p,\hat{m}^{\tau}\right)K\left(\hat x\right)\right)\\
	&\quad+\frac{1}{2}\sum_{\alpha,\beta}i\Gamma_{\alpha,\beta}^{(j)} 
	\left( \epsilon^{\beta\gamma\tau}\left(K\left(\hat p,\hat m_{j}^{\tau},\hat m_{j}^{\alpha}\right)+K\left(\hat p,\hat m_{j}^{\tau}\right)K\left(\hat m_{j}^{\alpha}\right)+K\left(\hat p,\hat m_{j}^{\alpha}\right)K\left(\hat m_{j}^{\tau}\right)\right)\right.\nonumber\\
	&\qquad\qquad\qquad\left.+\epsilon^{\gamma\alpha\tau}\left(K\left(\hat p,\hat m_{j}^{\beta},\hat m_{j}^{\tau}\right)+K\left(\hat p,\hat m_{j}^{\beta}\right)K\left(\hat m_{j}^{\tau}\right)+K\left(\hat p,\hat m_{j}^{\tau}\right)K\left(\hat m_{j}^{\beta}\right)\right)\right),\nonumber
\end{align}
\begin{align}
	\frac{d}{dt} K\left(\hat m_{j}^\gamma,\hat m_{\ell}^\zeta\right) 
	&=-\sum_{\alpha}\left(\epsilon^{\alpha\gamma\tau}\omega_{\alpha}^{(j)}K\left(\hat m_{j}^{\tau},\hat m_{\ell}^\zeta\right)+\epsilon^{\alpha\zeta\tau} \omega_{\alpha}^{(\ell)}K\left(\hat m_{j}^{\gamma}, \hat m_{\ell}^{\tau}\right)\right)\nonumber\\
	&\quad-\sum_{\alpha}g_{\alpha}\sqrt{2\omega_{b}}\left(\epsilon^{\alpha\gamma\tau}\left(K\left(\hat x,\hat m_{j}^{\tau},\hat m_{\ell}^\zeta\right)+K\left(\hat x,\hat m_{j}^{\tau}\right)K\left(\hat m_{\ell}^\zeta\right)+K\left(\hat m_{j}^{\tau},\hat m_{\ell}^\zeta\right)K\left(\hat x\right)\right)\right.\nonumber\\
	&\qquad\qquad\qquad\left.+\epsilon^{\alpha\zeta\tau} \left(K\left(\hat x,\hat m_{j}^{\gamma}, \hat m_{\ell}^{\tau}\right)+K\left(\hat x,\hat m_{j}^{\gamma}\right)K\left(\hat m_{\ell}^{\tau}\right)+K\left(\hat m_{j}^{\gamma}, \hat m_{\ell}^{\tau}\right)K\left(\hat x\right)\right)\right)\nonumber\\
	&\quad-\sum_{i=1}^M\sum_{\alpha,\beta}\left(\epsilon^{\beta\gamma\tau}\omega_{\alpha\beta}^{(i,j)}\left(K\left(\hat m_{i}^{\alpha},\hat m_{j}^{\tau},m_{\ell}^{\zeta}\right)+K\left(\hat m_{i}^{\alpha},m_{\ell}^{\zeta}\right)K\left(\hat m_{j}^{\tau}\right)+K\left(\hat m_{j}^{\tau},m_{\ell}^{\zeta}\right)K\left(\hat m_{i}^{\alpha}\right)\right)\right.\nonumber\\
	&\left.\qquad\qquad+\epsilon^{\alpha\gamma\tau}\omega_{\alpha\beta}^{(j,i)}\left(K\left(\hat m_{j}^{\tau},\hat m_{i}^{\beta},m_{\ell}^{\zeta}\right)+K\left(\hat m_{j}^{\tau},m_{\ell}^{\zeta}\right)K\left(\hat m_{i}^{\beta}\right)+K\left(\hat m_{i}^{\beta},m_{\ell}^{\zeta}\right)K\left(\hat m_{j}^{\tau}\right)\right)\right.\nonumber\\
	&\left.\qquad\qquad+\epsilon^{\beta\zeta\tau}\omega_{\alpha\beta}^{(i,k)}\left(K\left(\hat m_{j}^{\gamma},\hat m_{i}^{\alpha},m_{\ell}^{\tau}\right)+K\left(\hat m_{j}^{\gamma},\hat m_{i}^{\alpha}\right)K\left(\hat m_{\ell}^{\tau}\right)+K\left(\hat m_{j}^{\gamma},\hat m_{\ell}^{\tau}\right)K\left(\hat m_{i}^{\alpha}\right)\right)\right.\nonumber\\
	&\left.\qquad\qquad+\epsilon^{\alpha\zeta\tau}\omega_{\alpha\beta}^{(\ell,i)}\left(K\left(\hat m_{j}^{\gamma},\hat m_{\ell}^{\tau},\hat m_{i}^{\beta}\right)+K\left(\hat m_{j}^{\gamma},\hat m_{\ell}^{\tau}\right)K\left(\hat m_{i}^{\beta}\right)+K\left(\hat m_{j}^{\gamma},\hat m_{i}^{\beta}\right)K\left(\hat m_{\ell}^{\tau}\right)\right)\right)\nonumber\\
	&\quad+\frac{i}{2}\sum_{\alpha,\beta}\left(\Gamma_{\alpha,\beta}^{(j)}\epsilon^{\beta\gamma\tau}\left(K\left(\hat m_{j}^{\tau},\hat m_{\ell}^{\zeta},\hat m_{j}^{\alpha}\right)+K\left(\hat m_{j}^{\tau},\hat m_{\ell}^{\zeta}\right)K\left(\hat m_{j}^{\alpha}\right)+K\left(\hat m_{\ell}^{\zeta},\hat m_{j}^{\alpha}\right)K\left(\hat m_{j}^{\tau}\right)\right)\right.\nonumber\\
	&\left.\qquad\qquad+\Gamma_{\alpha,\beta}^{(j)}\epsilon^{\gamma\alpha\tau}\left(K\left(\hat m_{j}^{\beta},\hat m_{j}^{\tau},\hat m_{\ell}^{\zeta}\right)+K\left(\hat m_{j}^{\beta},\hat m_{\ell}^{\zeta}\right)K\left(\hat m_{j}^{\tau}\right)+K\left(\hat m_{j}^{\tau},\hat m_{\ell}^{\zeta}\right)K\left(\hat m_{j}^{\beta}\right)\right)\right.\nonumber\\
	&\left.\qquad\qquad+\Gamma_{\alpha,\beta}^{(\ell)}\epsilon^{\beta\zeta\tau}\left(K\left(\hat m_{j}^{\gamma},\hat m_{\ell}^{\tau},\hat m_{\ell}^{\alpha}\right)+K\left(\hat m_{j}^{\gamma},\hat m_{\ell}^{\tau}\right)K\left(\hat m_{\ell}^{\alpha}\right)+K\left(\hat m_{j}^{\gamma},\hat m_{\ell}^{\alpha}\right)K\left(\hat m_{\ell}^{\tau}\right)\right)\right.\nonumber\\
	&\left.\qquad\qquad+\Gamma_{\alpha,\beta}^{(\ell)}\epsilon^{\zeta\alpha\tau}\left(K\left(\hat m_{\ell}^{\beta},\hat m_{j}^{\gamma},\hat m_{\ell}^{\tau}\right)+K\left(\hat m_{\ell}^{\beta},\hat m_{j}^{\gamma}\right)K\left(\hat m_{\ell}^{\tau}\right)+K\left(\hat m_{j}^{\gamma},\hat m_{\ell}^{\tau}\right)K\left(\hat m_{\ell}^{\beta}\right)\right)\right).\nonumber
\end{align}

 We can see that an initial uncorrelated condition $\lim_{N\rightarrow\infty}K\left(\hat o_{j}(0),\hat o_{k}(0)\right)=0$ implies that $\frac{d}{dt} \lim_{N\rightarrow\infty}K\left(\hat o_{j},\hat o_{k}\right)= 0$, $\forall$ $j,k$. 
Given that the dynamical equations are well defined/convergent in the thermodynamic (such as in the simplest case of couplings $\omega_{[...]}, g_\alpha, \kappa$ and $\Gamma$ as finite constants), such a null time derivative implies that the system second cumulants remain null for an infinitesimal evolution. In other words, the system remains uncorrelated therefore still constrained to only MF first order cumulants. We can directly iterate this reasoning throughout the entire dynamics of the system, for which at any time no correlations are generated, thus proving the exactness of MF.
As an illustration we may choose an initial state composed of the tensor product of a single-mode coherent state with a separable pure spin state which leads to cumulants of the form,
\begin{equation}
	K\left(\hat o_{j}(0),\hat o_{k}(0)\right)=\frac{c_{j,k}}{N}, \quad K\left(\hat o_{j}(0),\hat o_{k}(0),\hat o_{\ell}(0)\right)=\frac{c_{j,k,\ell}}{N^2},
\end{equation}
where $c_{j,k}, c_{j,k,\ell}\in\mathbb{C}$. Notice that also higher order cumulants of the uncorrelated state shall vanish with $N$ in a similar form.

In summary, the MF semiclassical approach is exact as long as the initial state is uncorrelated according to Eq.\eqref{eq.uncorrelatedMF.state} in the main text, the dynamical equations are well defined (convergent) in the thermodynamic limit and since they describe macroscopic observable dynamics, the corresponding dynamical rates in the Heisenberg equations of motion must be extensive with the system size.  If the dynamical rate is sub-extensive for some state observables it can lead to an unstable fixed point in the MF dynamics, not related though to a steady state rather a fine-tuning point. A simple example can be borrowed from the system with a single spin ensemble and purely dissipative decay , $\Gamma_{x,x} = \Gamma_{y,y} = 1$, $ \Gamma_{y,x} = i = \Gamma_{x,y}^*$ with other parameters null. The product state with all spins up has a  sub-extensive  dynamical rate in the Heisenberg equation of motion, therefore it is a fixed point in MF equations despite simply an unstable fine-tuning  point.

\section*{Pair of collective spin-1/2 systems}

\begin{figure}
		\includegraphics[width = 0.49 \linewidth]{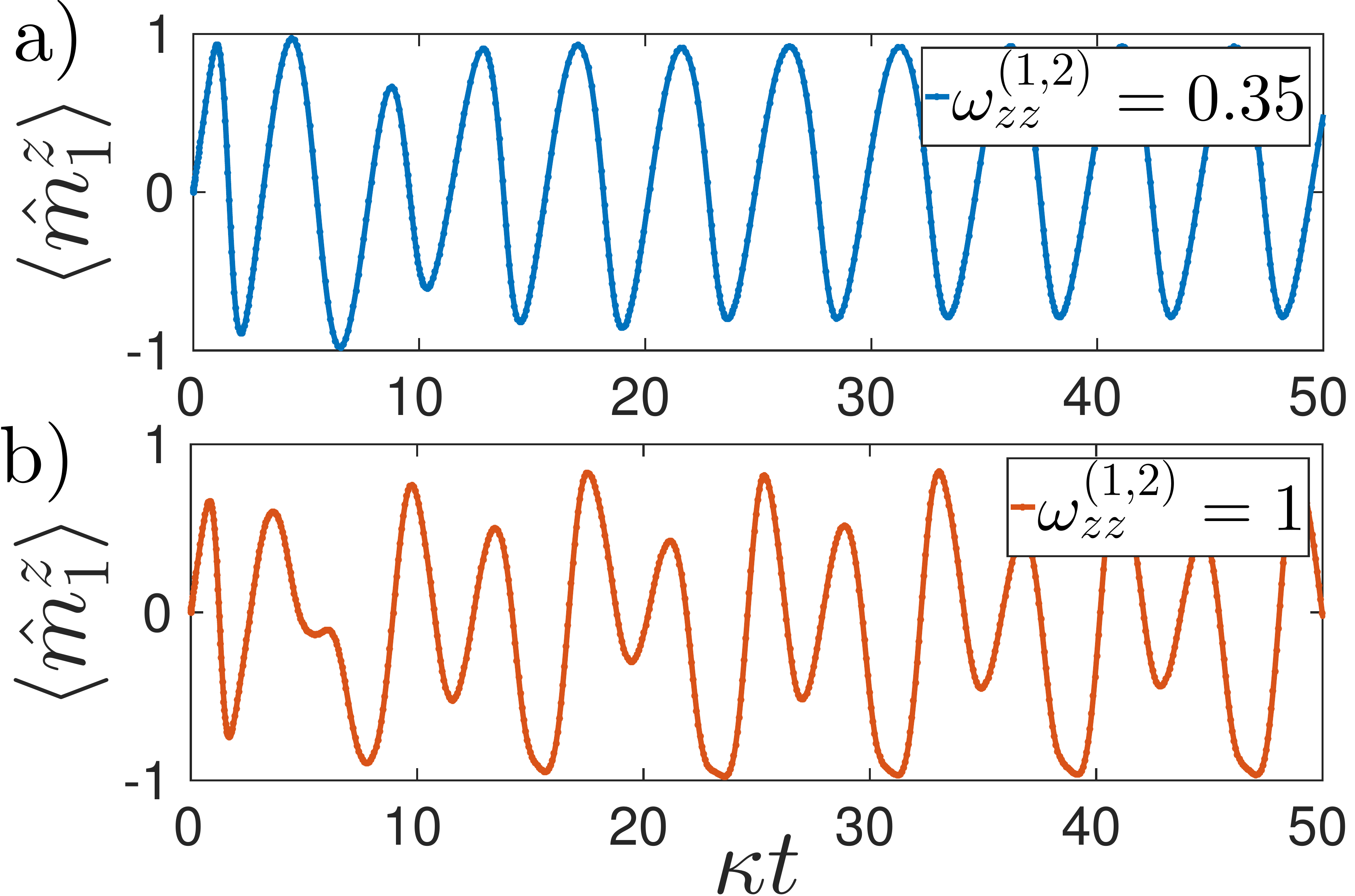}
		\includegraphics[width = 0.49 \linewidth]{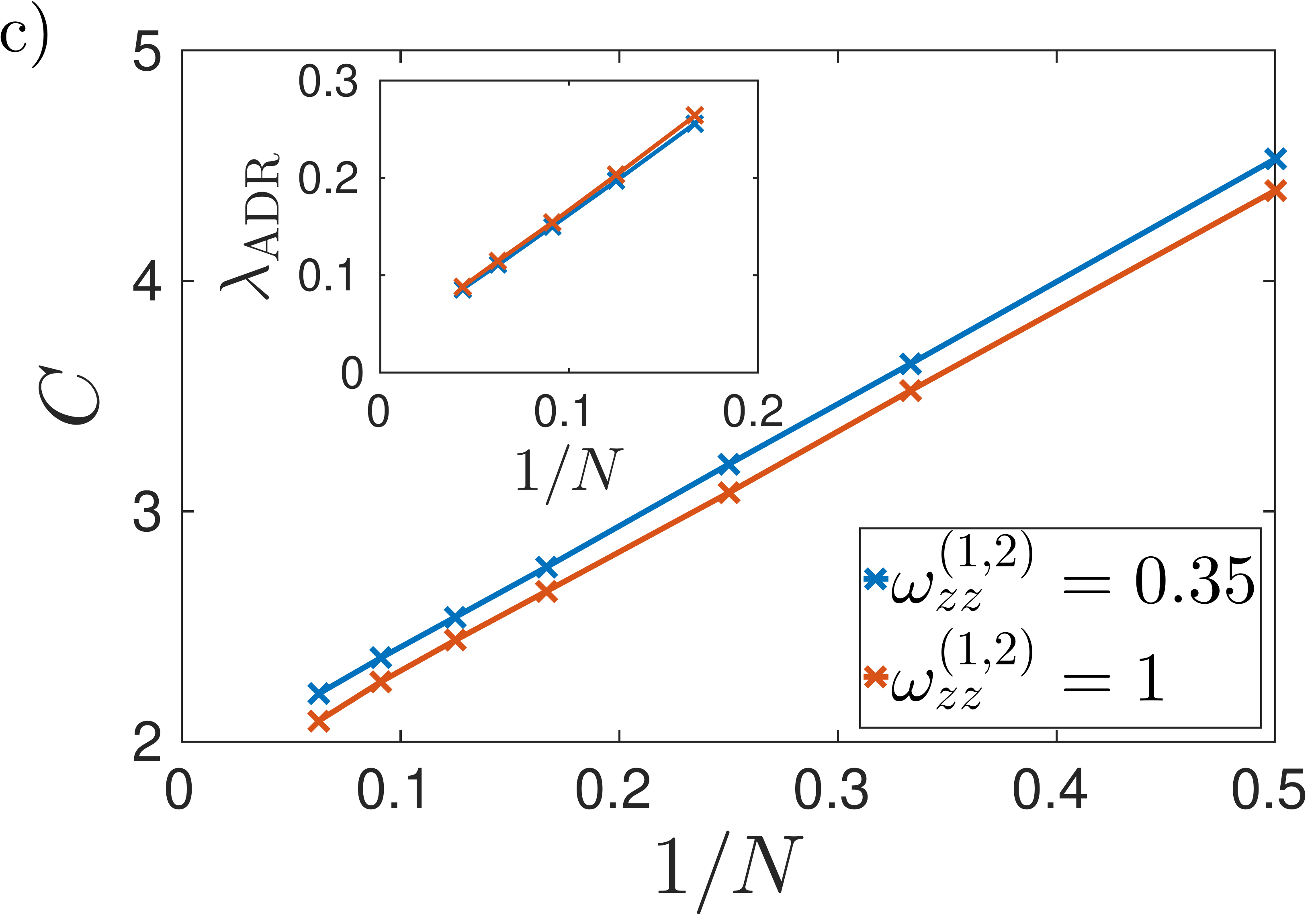}
		\caption{
		Dynamics for a pair of collective spin $1/2$ systems in the thermodynamic limit, along its (a) limit-cycle and (b) period doubling dynamics, with $\omega_{zz} = 0.35 $ and $1$, respectively. In both cases we set $\omega_{xx}^{(1,2)} = 3$, $\omega_{z}^{(1)} = 0.1$, $\omega_{z}^{(2)} = 0.02$ and $\omega_x^{(j)}/\kappa = 2$ for $j=1,2$ . \textbf{(c)} Finite-size scaling for the cumulant correlations $C = \sum_{j \leq \ell} \sum_{\alpha,\beta} K(\hat m_j^\alpha,\hat m_\ell^\beta)$ and (inset panel) asymptotic decay rate $\lambda_{\rm ADR}$.
	}
	\label{fig:SU22}
\end{figure}

 Interacting collective systems can host more general forms of dynamics~\cite{Luis2021}. We consider a pair of interacting spin ensembles ($M=2$). Both ensembles are locally driven similarly to the single collective spin-$1/2$ system, with $\Gamma_{x,x}^{(j)} = \Gamma_{y,y}^{(j)} = \sqrt{\kappa}$, $ \Gamma_{y,x}^{(j)} = i \sqrt{\kappa} = (\Gamma_{x,y}^{(j)})^*$ and  $\omega_{x}^{(j)}/\kappa = 2$ for $j=1,2$, thus in the weak dissipative regime.  Once coupled, they may synchronize their motion leading to richer dynamics, as we observe in Fig.(\ref{fig:SU22}a-b) for nonnull $\omega_{xx}$ and $\omega_{zz}$ interactions. The global dynamics in the thermodynamic limit synchronizes towards limit-cycles (Fig.(\ref{fig:SU22})) where after an initial transient relaxation time the spins lock their motion to a single frequency (Fig.(\ref{fig:SU22}a) or period doubling bifurcations (Fig.(\ref{fig:SU22}b). Performing a finite-size scaling analysis for the NESS correlations we obtain in both cases that cumulant correlations are nonzero, as shown in Fig.(\ref{fig:SU22}c). Moreover, one can indirectly extract the Lindbaldian gap from an asymptotic decay rate analysis, which we denote as $\lambda_{\rm ADR}$. We observe the presence of gapless excitations (inset of Fig.(\ref{fig:SU22}c) for such interacting time-crystal phases, corroborating our expectation.

The asymptotic decay rate analysis if performed as follows. In the long time limit, the Lindbladian dynamics is dominated only by its slowest decay modes.  Given the dynamics of an observable $o$, its long time limit can be approximated by
	\begin{equation}
		o(t) - o(t \rightarrow \infty) \sim  e^{\lambda_{\rm ADR }t }
	\end{equation}
	with $\lambda_{\rm ADR }$ the asymptotic decay rate. Usually such decay rate is directly related to the gap of the Lindbladian; nevertheless, in the general case it corresponds to a upper bound for the gap.

\begin{figure}
	\includegraphics[scale = 0.3]{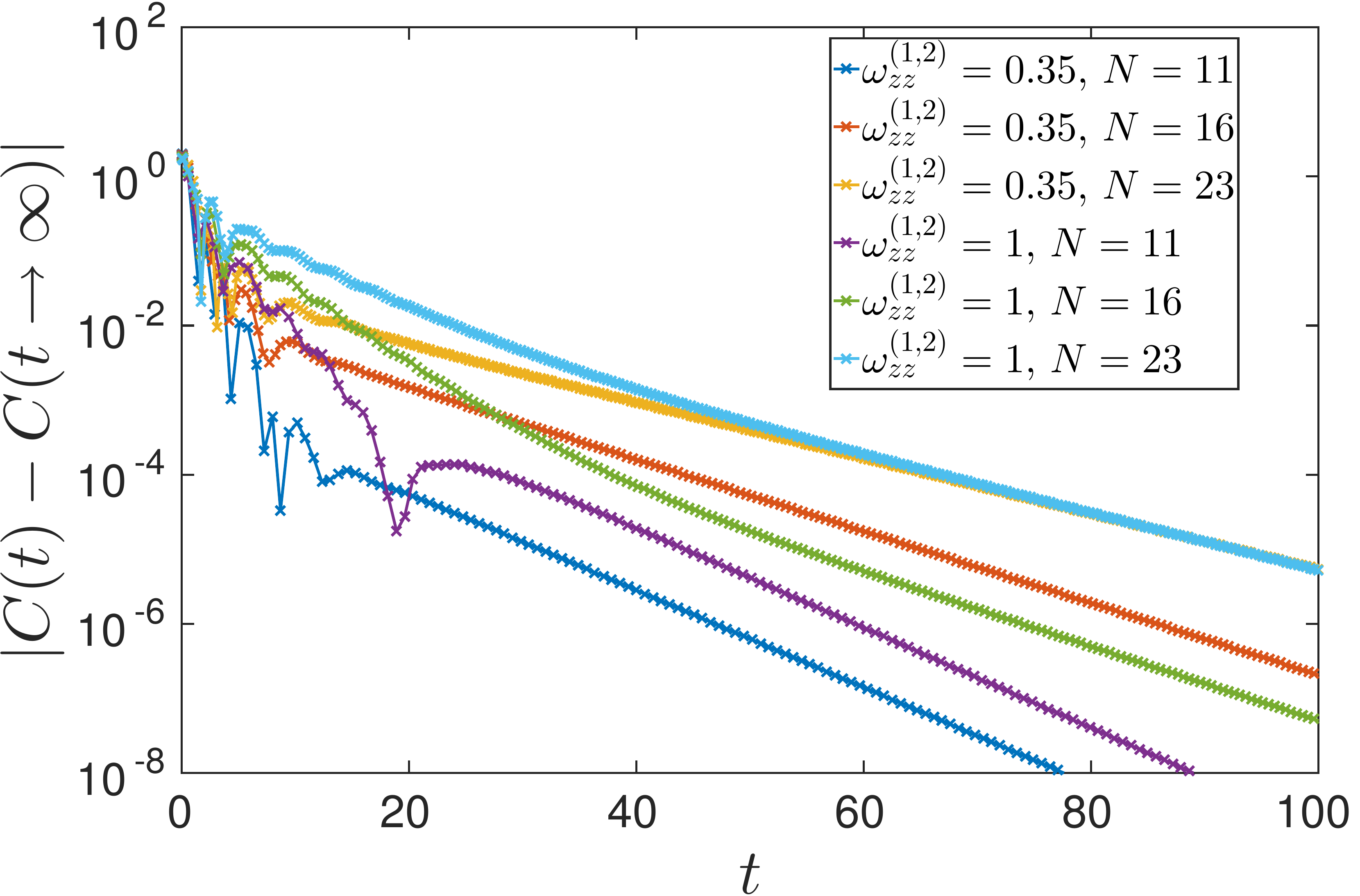}
	\caption{Dynamics of cumulant correlations $C=\sum_{j \leq \ell} \sum_{\alpha,\beta} K(\hat m_j^\alpha,\hat m_\ell^\beta)$ approaching its steady state value in the pair collective spin-$1/2$ system. We see the exponential decay in the long time limit, for which we can infer the asymptotic decay rate of the dynamics.}
	\label{fig:SI_dyn_su22}
\end{figure}

We extract the asymptotic decay rate for the pair of collective spin-$1/2$ systems. We show in Fig.\eqref{fig:SI_dyn_su22} the asymptotic dynamics for the cumulant correlations. We see that for long time it approaches an exponential decay, for which one can numerically extract its decay rate. We see that for larger systems size $\lambda_{\rm ADR}$ decreases, highlighting its gapless nature.

\section*{Hermitian Lindblad operators}

We show here further details of the spectral gap and dynamics for the Lindbladian with Hermitian jump operators discussed in the main text.
The model corresponds to a single spin ensemble ($M=1$) driven by a coherent field Hamiltonian along the $x$-direction ($\omega_x$) and a dissipation along the orthogonal $z$-direction ($\Gamma_{z,z}$), with all other parameters null in the Lindbladian.

\begin{figure}
	\includegraphics[width = 0.45 \linewidth]{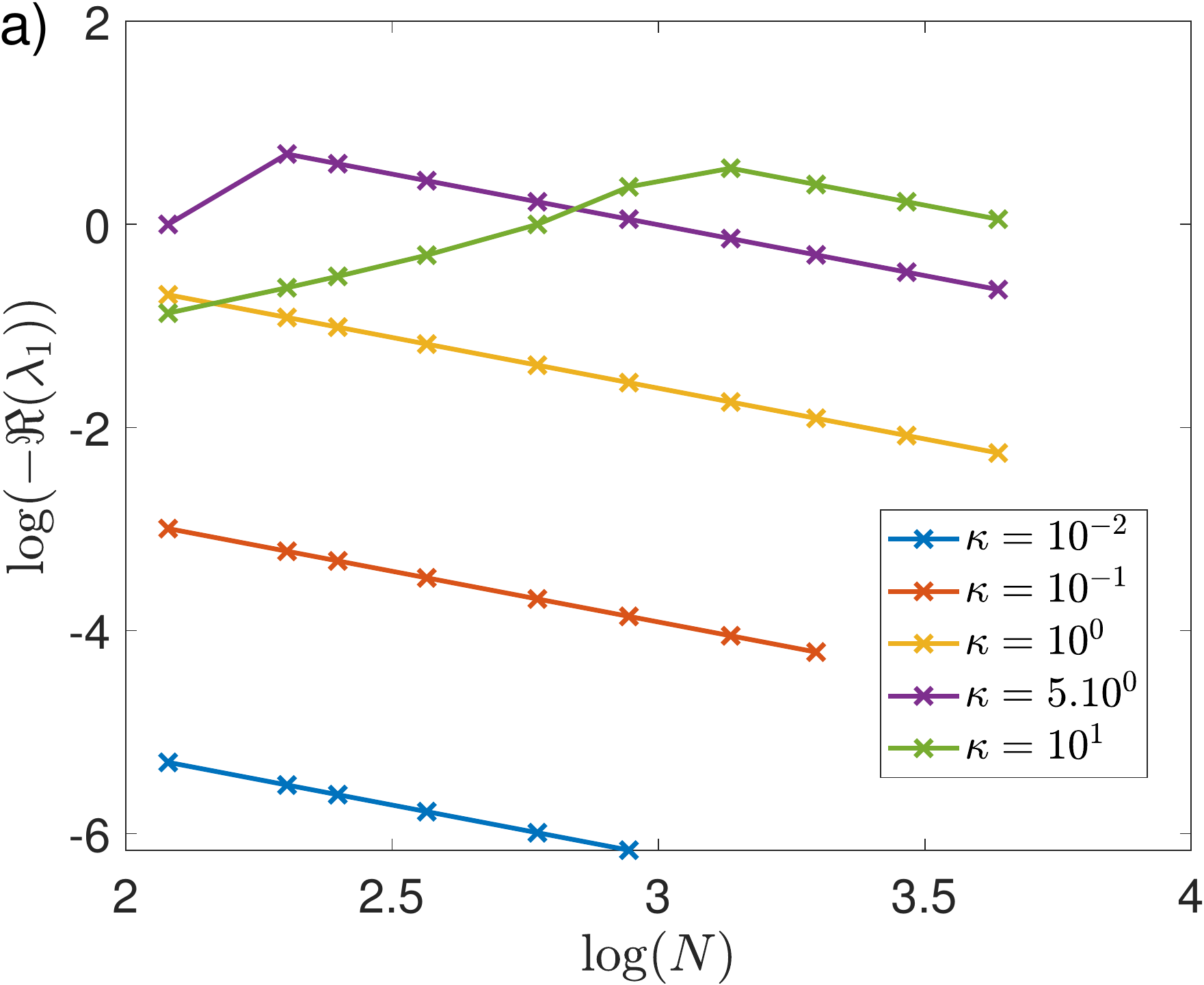}
	\includegraphics[width = 0.45 \linewidth]{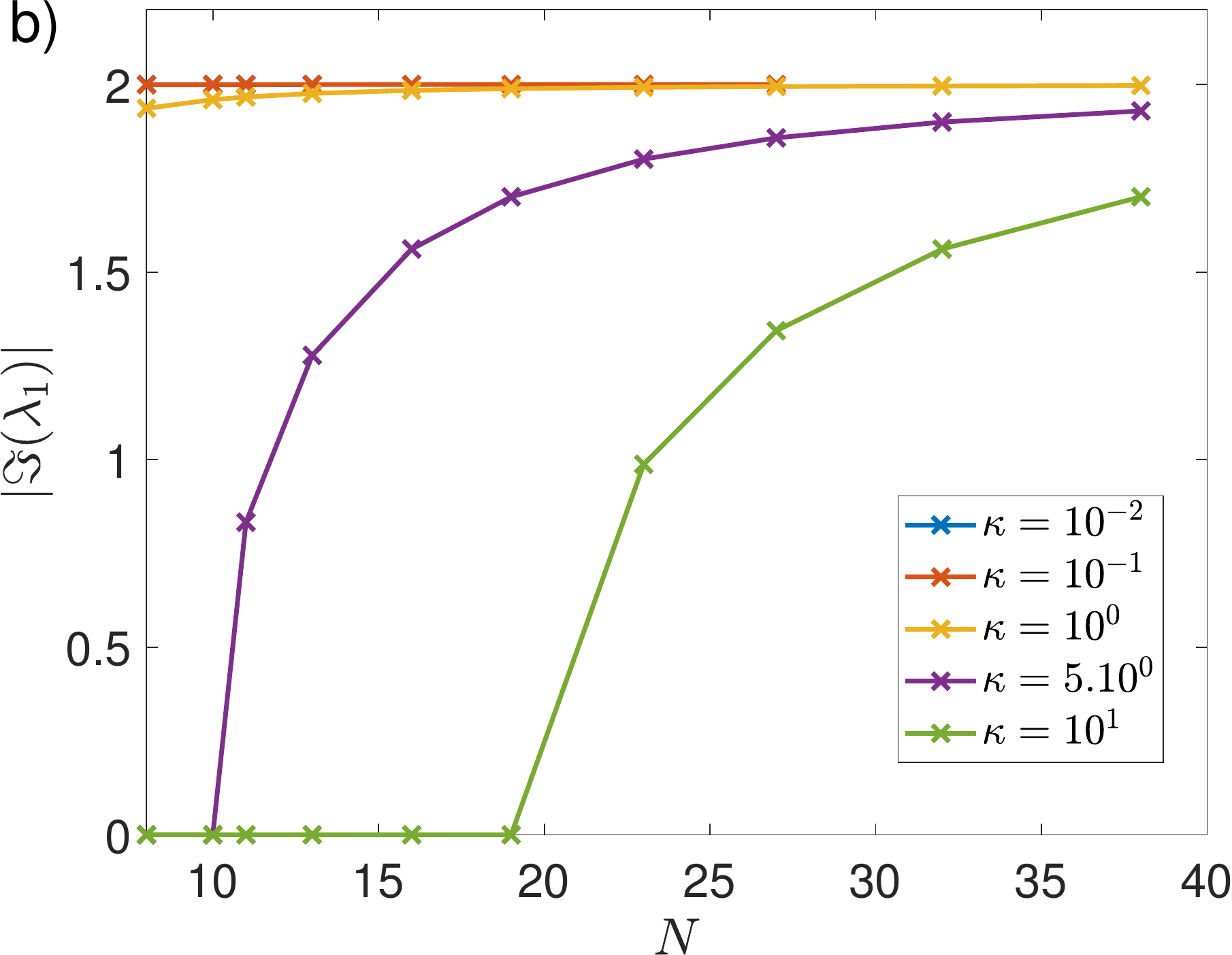}
	\includegraphics[width = 0.45 \linewidth]{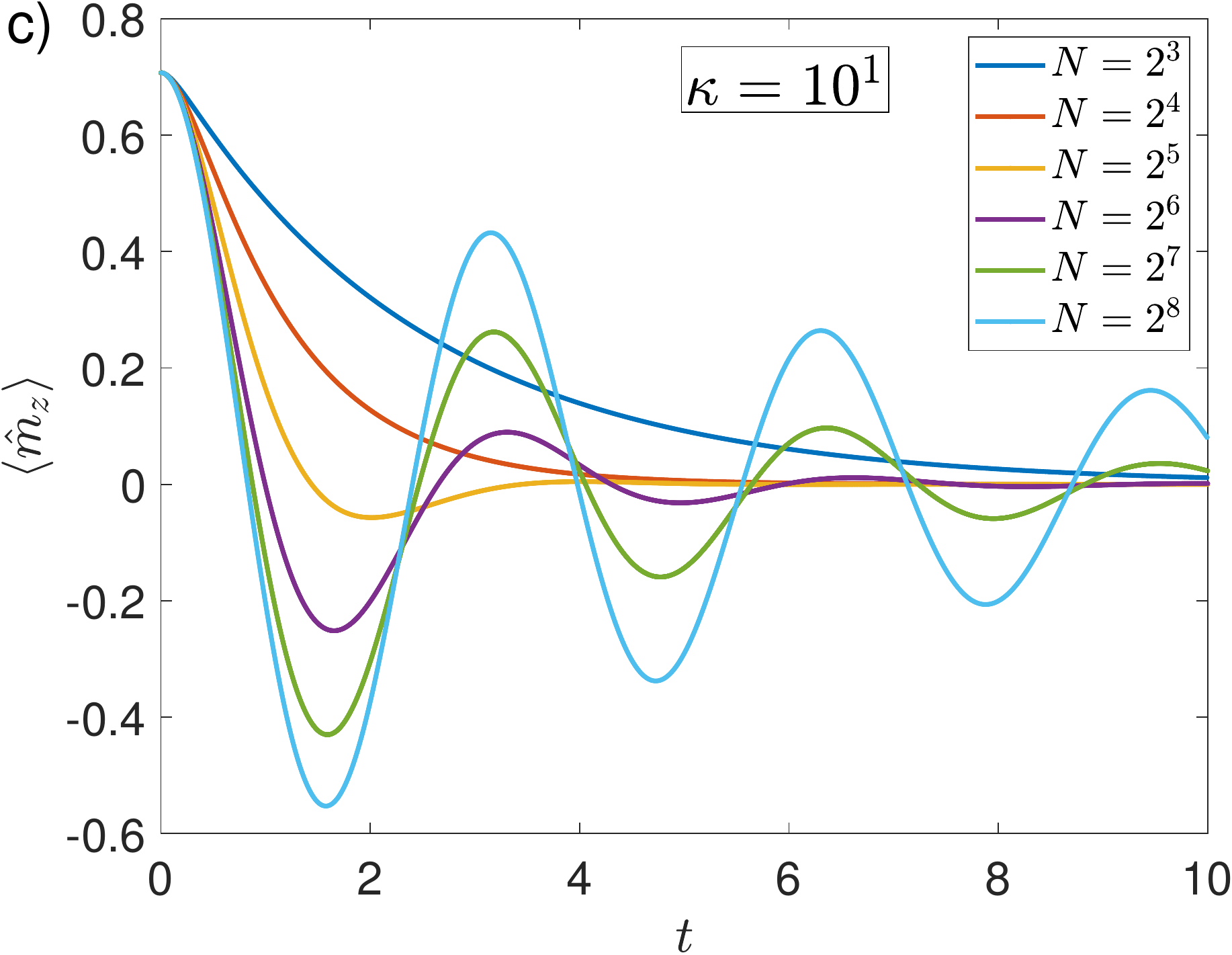}
	\includegraphics[width = 0.45 \linewidth]{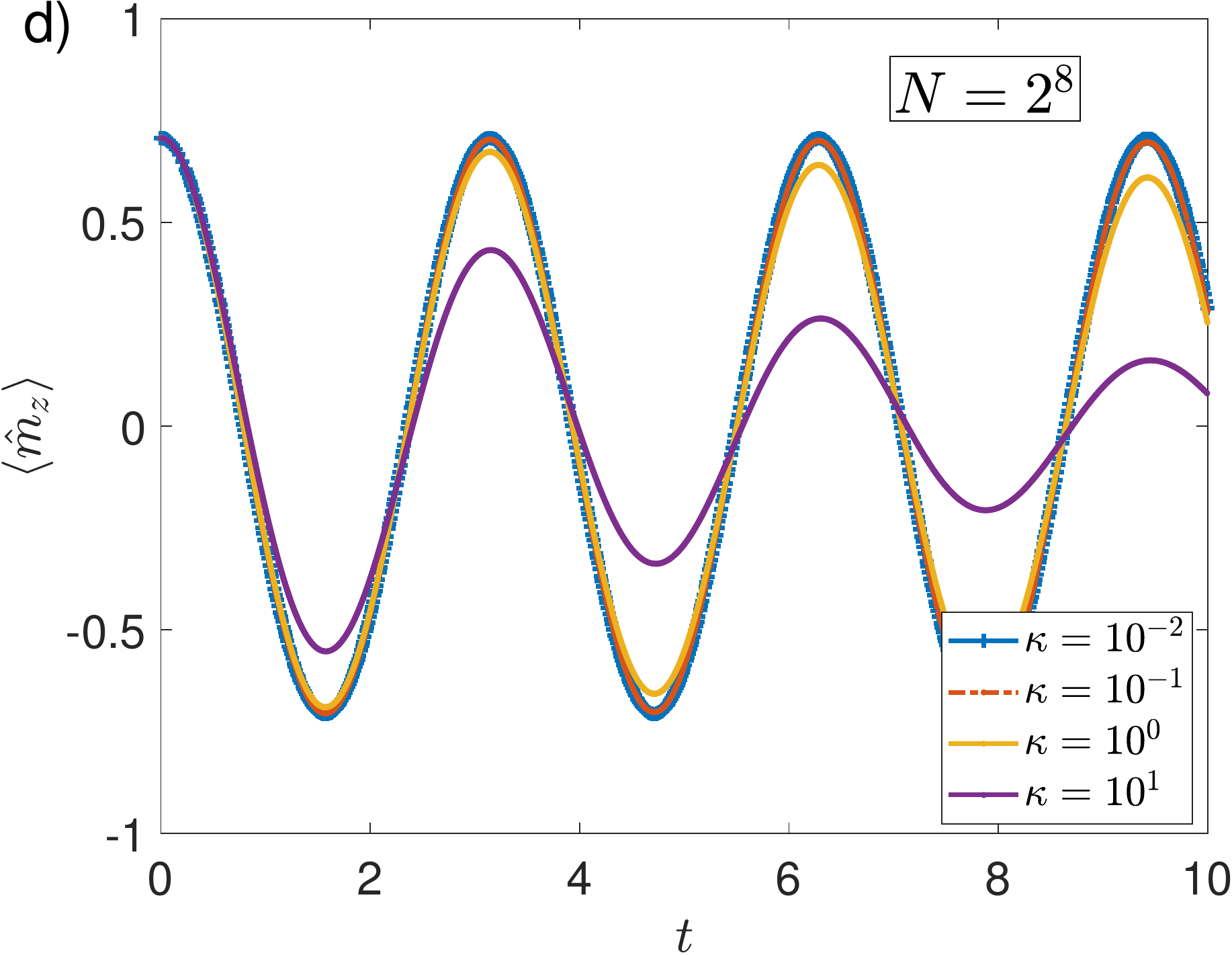}
	\caption{  \textbf{(Upper panels)} We show the (a) real and (b) imaginary part of the gap eigenvalue, for varying system sizes and coupling strength. 
		\textbf{(Bottom panels)} Dynamics for the macroscopic magnetization along $z$-direction, given an initial coherent spin state
		$|\psi(0)\rangle = e^{-i\frac{\pi}{4} \hat m_y} |\uparrow ... \uparrow\rangle$
		. We show in the (c) the dynamics for different system sizes, in a strong dissipative regime with $\kappa/\omega_x=10$. In the (d) we show the dynamics for varying couplings in a system with $N=2^8$ spins. In all plots we set $\omega_x=1$.
	}
	\label{fig:hermitian.jump}
\end{figure}

In Fig.\eqref{fig:hermitian.jump}-(upper panels) we show the real and imaginary part of the gap eigenvalue for different system sizes and coupling strength. For all coupling strengths studied, wee see that the real part of the eigenvalue features a $1/N$ decay rate for sufficiently large system sizes, accompanied by an imaginary term approaching $2\omega_x$. In Fig.\eqref{fig:hermitian.jump}-(bottom  panels) we show the dynamics for the macroscopic magnetization given an initial coherent state 
$|\psi(0)\rangle = e^{-i\frac{\pi}{4} \hat m_y} |\uparrow ... \uparrow\rangle$ in the evolution. We see that the life-time for the oscillations diverge with system size, leading to persistent dynamics in the thermodynamic limit.

\section*{Exactness of semiclassical permutation invariant dynamics}

In this section we show further details of the exactness of the mean-field approach in the thermodynamic limit for the permutationally invariant model discussed in the main text. We consider the purely dissipative channel for a single spin ensemble $(M = 1)$, for simplicity. Therefore in our notation we omit the spin ensemble index for the spin operators, 
$\hat \sigma_{i,k} \equiv \hat \sigma_k$.
The dissipative channel is described by,
\begin{align}\label{eq.master.equation.inv.perm}
	\mathcal{D}[ \hat\rho]
	&=\sum_{j,k=1}^N\sum_{\alpha,\beta}\gamma_{\alpha,\beta}^{j,k} \left(\hat \sigma_{j}^{\alpha} \hat\rho  \hat \sigma_{k}^{\beta} - \frac{1}{2}\{\hat \sigma_{k}^{\beta} \hat \sigma_{j}^{\alpha},\hat\rho \}\right),
\end{align}
where  $\gamma_{\alpha,\beta}^{j,k}\in\mathbb{C}$ are the elements of the dissipative spin matrix.
The dynamics of a general operator $\hat o$ within the Heisenberg picture can be written as, 
\begin{equation}
	\frac{ d \langle \hat o \rangle }{dt} = \frac{1}{2}\sum_{j,k=1}^N\sum_{\alpha,\beta}\gamma_{\alpha,\beta}^{j,k} 
	\langle [\hat \sigma_{k}^{\beta},\hat o ] \hat \sigma_{j}^{\alpha}
	+ \hat \sigma_{k}^{\beta}[\hat o ,\hat \sigma_{j}^{\alpha}] \rangle.
	\label{eq:dynamical.equation.chain}
\end{equation}
Considering the macroscopic operators $\hat m_\alpha =\hat S_\alpha/S$ and using SU(2) commutation relations, $[\hat \sigma_{j}^\alpha,\hat \sigma_{k}^\beta] = 2i\epsilon^{\alpha \beta \gamma}\hat \sigma_{j}^\gamma\delta_{jk}$, we can obtain the corresponding  one-body and two-body equations of motion,
\begin{align}
	\frac{d}{dt} \langle\hat m_\zeta\rangle  
	&=\frac{i}{N}\sum_{\alpha,\beta}\sum_{j=1}^N\sum_{k\neq j}\gamma_{\alpha,\beta}^{j,k} 
	\left( \epsilon^{\beta\zeta\tau}\langle\hat\sigma_{k}^{\tau}\hat\sigma_{j}^{\alpha}\rangle
	+ \epsilon^{\zeta\alpha\tau}\langle\hat\sigma_{k}^{\beta}\hat\sigma_{j}^{\tau}\rangle\right)\nonumber\\
	&\quad+\frac{i}{N}\sum_{\alpha,\beta}\sum_{j=1}^N\gamma_{\alpha,\beta}^{j,j} 
	\left( \epsilon^{\beta\zeta\tau}\left(\delta_{\tau\alpha}+i\epsilon^{\tau\alpha\tau^{\prime}}\langle\hat\sigma_{j}^{\tau^{\prime}}\rangle\right)
	+ \epsilon^{\zeta\alpha\tau}\left(\delta_{\beta\tau}+i\epsilon^{\beta\tau\tau^{\prime}}\langle\hat\sigma_{j}^{\tau^{\prime}}\rangle\right)\right),\nonumber
\end{align}
\begin{align}
	\frac{d}{dt} \langle\hat m_\zeta\hat m_\eta\rangle 
	&=\frac{i}{N^2}\sum_{\alpha,\beta}\sum_{j=1}^N\left(\sum_{k\neq j}\sum_{\ell\neq k,\ell\neq j}\gamma_{\alpha,\beta}^{j,k}\left(\epsilon^{\beta\zeta\tau}\langle\hat \sigma_{k}^{\tau}\hat \sigma_{\ell}^{\eta}\hat \sigma_{j}^{\alpha}\rangle+\epsilon^{\beta\eta\tau}\langle\hat \sigma_{\ell}^{\zeta}\hat \sigma_{k}^{\tau}\hat \sigma_{j}^{\alpha}\rangle+\epsilon^{\zeta\alpha\tau}\langle\hat \sigma_{k}^{\beta}\hat \sigma_{j}^{\tau}\hat \sigma_{\ell}^{\eta}\rangle+\epsilon^{\eta\alpha\tau}\langle\hat \sigma_{k}^{\beta}\hat \sigma_{\ell}^{\zeta}\hat \sigma_{j}^{\tau}\rangle\right)\right.\nonumber\\
	&\quad\qquad\qquad\qquad\left.+\sum_{k\neq j}\gamma_{\alpha,\beta}^{j,k}\left(\epsilon^{\beta\zeta\tau}\left(\delta_{\eta\alpha}\langle\hat \sigma_{k}^{\tau}\rangle+\delta_{\tau\eta}\langle\hat \sigma_{j}^{\alpha}\rangle+i\epsilon^{\eta\alpha\tau^{\prime}}\langle\hat \sigma_{k}^{\tau}\hat \sigma_{j}^{\tau^{\prime}}\rangle+i\epsilon^{\tau\eta\tau^{\prime}}\langle\hat \sigma_{k}^{\tau^\prime}\hat \sigma_{j}^{\alpha}\rangle\right)\right.\right.\nonumber\\
	&\qquad\qquad\qquad\qquad\qquad\qquad\left.\left.+\epsilon^{\beta\eta\tau}\left(\delta_{\zeta\alpha}\langle\hat \sigma_{k}^{\tau}\rangle+\delta_{\zeta\tau}\langle\hat \sigma_{j}^{\alpha}\rangle+i\epsilon^{\zeta\alpha\tau^{\prime}}\langle\hat \sigma_{k}^{\tau}\hat \sigma_{j}^{\tau^{\prime}}\rangle+i\epsilon^{\zeta\tau\tau^{\prime}}\langle\hat \sigma_{k}^{\tau^\prime}\hat \sigma_{j}^{\alpha}\rangle\right)\right.\right.\nonumber\\
	&\qquad\qquad\qquad\qquad\qquad\qquad\left.\left.+\epsilon^{\zeta\alpha\tau}\left(\delta_{\tau\eta}\langle\hat \sigma_{k}^{\beta}\rangle+\delta_{\beta\eta}\langle\hat \sigma_{j}^{\tau}\rangle+i\epsilon^{\tau\eta\tau^{\prime}}\langle\hat \sigma_{k}^{\beta}\hat \sigma_{j}^{\tau^{\prime}}\rangle+i\epsilon^{\beta\eta\tau^{\prime}}\langle\hat \sigma_{k}^{\tau^\prime}\hat \sigma_{j}^{\tau}\rangle\right)\right.\right.\nonumber\\
	&\qquad\qquad\qquad\qquad\qquad\qquad\left.\left.+\epsilon^{\eta\alpha\tau}\left(\delta_{\zeta\tau}\langle\hat \sigma_{k}^{\beta}\rangle+\delta_{\beta\zeta}\langle\hat \sigma_{j}^{\tau}\rangle+i\epsilon^{\zeta\tau\tau^{\prime}}\langle\hat \sigma_{k}^{\beta}\hat \sigma_{j}^{\tau^{\prime}}\rangle+i\epsilon^{\beta\zeta\tau^{\prime}}\langle\hat \sigma_{k}^{\tau^\prime}\hat \sigma_{j}^{\tau}\rangle\right)\right)\right.\\
	&\qquad\qquad\qquad\quad\left.+\sum_{\ell\neq j}\gamma_{\alpha,\beta}^{j,j}\left(\epsilon^{\beta\zeta\tau}\left(\delta_{\tau\alpha}\langle\hat \sigma_{\ell}^{\eta}\rangle+i\epsilon^{\tau\alpha\tau^{\prime}}\langle\hat \sigma_{\ell}^{\eta}\hat \sigma_{j}^{\tau^{\prime}}\rangle\right)+\epsilon^{\beta\eta\tau}\left(\delta_{\tau\alpha}\langle\hat \sigma_{\ell}^{\zeta}\rangle+i\epsilon^{\tau\alpha\tau^{\prime}}\langle\hat \sigma_{\ell}^{\zeta}\hat \sigma_{j}^{\tau^{\prime}}\rangle\right)\right.\right.\nonumber\\
	&\qquad\qquad\qquad\qquad\qquad\qquad\left.\left.+\epsilon^{\zeta\alpha\tau}\left(\delta_{\beta\tau}\langle\hat \sigma_{\ell}^{\eta}\rangle+i\epsilon^{\beta\tau\tau^{\prime}}\langle\hat \sigma_{\ell}^{\eta}\hat \sigma_{j}^{\tau^{\prime}}\rangle\right)+\epsilon^{\eta\alpha\tau}\left(\delta_{\beta\tau}\langle\hat \sigma_{\ell}^{\zeta}\rangle+i\epsilon^{\beta\tau\tau^{\prime}}\langle\hat \sigma_{\ell}^{\zeta}\hat \sigma_{j}^{\tau^{\prime}}\rangle\right)\right)\right.\nonumber\\
	&\qquad\qquad\qquad\qquad+\left.\gamma_{\alpha,\beta}^{j,j}\left(\epsilon^{\beta\zeta\tau}\left(\delta_{\tau\eta}\langle\hat \sigma_{j}^{\alpha}\rangle+i\epsilon^{\tau\eta\tau^{\prime}}\left(\delta_{\tau^\prime\alpha}+i\epsilon^{\tau^\prime\alpha\tau^{\prime\prime}}\langle\hat \sigma_{j}^{\tau^{\prime\prime}}\rangle\right)\right)\right.\right.\nonumber\\
	&\qquad\qquad\qquad\qquad\qquad\qquad\left.\left.+\epsilon^{\beta\eta\tau}\left(\delta_{\zeta\tau}\langle\hat \sigma_{j}^{\alpha}\rangle+i\epsilon^{\zeta\tau\tau^{\prime}}\left(\delta_{\tau^\prime\alpha}+i\epsilon^{\tau^\prime\alpha\tau^{\prime\prime}}\langle\hat \sigma_{j}^{\tau^{\prime\prime}}\rangle\right)\right)\right.\right.\nonumber\\
	&\qquad\qquad\qquad\qquad\qquad\qquad\left.\left.+\epsilon^{\zeta\alpha\tau}\left(\delta_{\tau\eta}\langle\hat \sigma_{j}^{\beta}\rangle+i\epsilon^{\tau\eta\tau^{\prime}}\left(\delta_{\beta\tau^\prime}+i\epsilon^{\beta\tau^\prime\tau^{\prime\prime}}\langle\hat \sigma_{j}^{\tau^{\prime\prime}}\rangle\right)\right)\right.\right.\nonumber\\
	&\qquad\qquad\qquad\qquad\qquad\qquad\left.\left.+\epsilon^{\eta\alpha\tau}\left(\delta_{\zeta\tau}\langle\hat \sigma_{j}^{\beta}\rangle+i\epsilon^{\zeta\tau\tau^{\prime}}\left(\delta_{\beta\tau^\prime}+i\epsilon^{\beta\tau^\prime\tau^{\prime\prime}}\langle\hat \sigma_{j}^{\tau^{\prime\prime}}\rangle\right)\right)\right)\right),\nonumber
\end{align}
where we have used the contraction equality $\hat \sigma_j^\alpha \sigma_j^\beta = \delta_{\alpha \beta}\mathbb{I} + i \epsilon^{\alpha \beta \tau} \hat \sigma_j^\tau$ for operators of the same spin. In the above equations we split the sums in $j,k,\ell$ and $m$ on all its possible combinations, and for the cases with equal indexes we contract the spin operators using the contracting equality.
Taking the initial state invariant under permutation, then the mean values $\langle\hat \sigma_{i}^{\alpha}\hat \sigma_{j}^{\beta}\dots\rangle$ doesn't depend on the spin lattice indices since the dynamics preserver the permutation symmetry, thus we can assume $\langle\hat \sigma_{i}^{\alpha}\hat \sigma_{j}^{\beta} \dots\rangle\equiv\langle\sigma^{\alpha}\hat \sigma^{\beta}\dots\rangle$. Using these definitions we can write,
\begin{align}
	\frac{d}{dt} \langle\hat m_\zeta\rangle  
	&=\sum_{\alpha,\beta}\Gamma_{\alpha,\beta}^{\rm off}(N)\left( \epsilon^{\beta\zeta\tau}\langle\sigma^{\tau}\hat\sigma^{\alpha}\rangle
	+ \epsilon^{\zeta\alpha\tau}\langle\hat\sigma^{\beta} \hat\sigma^{\tau}\rangle\right)\nonumber\\
	&\quad+\sum_{\alpha,\beta}\Gamma_{\alpha,\beta}^{\rm diag}(N)
	\left( \epsilon^{\beta\zeta\tau}\left(\delta_{\tau\alpha}+i\epsilon^{\tau\alpha\tau^{\prime}}\langle\hat\sigma^{\tau^{\prime}}\rangle\right)
	+ \epsilon^{\zeta\alpha\tau}\left(\delta_{\beta\tau}+i\epsilon^{\beta\tau\tau^{\prime}}\langle\hat\sigma^{\tau^{\prime}}\rangle\right)\right),\nonumber
\end{align}
\begin{align}
	\frac{d}{dt} \langle\hat m_\zeta\hat m_\eta\rangle 
	&=\frac{(N-2)}{N}\sum_{\alpha,\beta}\Gamma_{\alpha,\beta}^{\rm off}(N)\left(\epsilon^{\beta\zeta\tau}\langle\hat \sigma^{\tau}\hat \sigma^{\eta}\hat \sigma^{\alpha}\rangle+\epsilon^{\beta\eta\tau}\langle\hat \sigma^{\zeta}\hat \sigma^{\tau}\hat \sigma^{\alpha}\rangle+\epsilon^{\zeta\alpha\tau}\langle\hat \sigma^{\beta}\hat \sigma^{\tau}\hat \sigma^{\eta}\rangle+\epsilon^{\eta\alpha\tau}\langle\hat \sigma^{\beta}\hat \sigma^{\zeta}\hat \sigma^{\tau}\rangle\right)\nonumber\\
	&\quad+\frac{1}{N}\sum_{\alpha,\beta}\Gamma_{\alpha,\beta}^{\rm off}(N)\left(\epsilon^{\beta\zeta\tau}\left(\delta_{\eta\alpha}\langle\hat \sigma^{\tau}\rangle+\delta_{\tau\eta}\langle\hat \sigma^{\alpha}\rangle+i\epsilon^{\eta\alpha\tau^{\prime}}\langle\hat \sigma^{\tau}\hat \sigma^{\tau^{\prime}}\rangle+i\epsilon^{\tau\eta\tau^{\prime}}\langle\hat \sigma^{\tau^\prime}\hat \sigma^{\alpha}\rangle\right)\right.\nonumber\\
	&\left.\qquad\qquad\qquad+\epsilon^{\beta\eta\tau}\left(\delta_{\zeta\alpha}\langle\hat \sigma^{\tau}\rangle+\delta_{\zeta\tau}\langle\hat \sigma^{\alpha}\rangle+i\epsilon^{\zeta\alpha\tau^{\prime}}\langle\hat \sigma^{\tau}\hat \sigma^{\tau^{\prime}}\rangle+i\epsilon^{\zeta\tau\tau^{\prime}}\langle\hat \sigma^{\tau^\prime}\hat \sigma^{\alpha}\rangle\right)\right.\nonumber\\
	&\left.\qquad\qquad\qquad+\epsilon^{\zeta\alpha\tau}\left(\delta_{\tau\eta}\langle\hat \sigma^{\beta}\rangle+\delta_{\beta\eta}\langle\hat \sigma^{\tau}\rangle+i\epsilon^{\tau\eta\tau^{\prime}}\langle\hat \sigma^{\beta}\hat \sigma^{\tau^{\prime}}\rangle+i\epsilon^{\beta\eta\tau^{\prime}}\langle\hat \sigma^{\tau^\prime}\hat \sigma^{\tau}\rangle\right)\right.\nonumber\\
	&\left.\qquad\qquad\qquad+\epsilon^{\eta\alpha\tau}\left(\delta_{\zeta\tau}\langle\hat \sigma^{\beta}\rangle+\delta_{\beta\zeta}\langle\hat \sigma^{\tau}\rangle+i\epsilon^{\zeta\tau\tau^{\prime}}\langle\hat \sigma^{\beta}\hat \sigma^{\tau^{\prime}}\rangle+i\epsilon^{\beta\zeta\tau^{\prime}}\langle\hat \sigma^{\tau^\prime}\hat \sigma^{\tau}\rangle\right)\right)\\
	&\quad+\frac{(N-1)}{N}\sum_{\alpha,\beta}\Gamma_{\alpha,\beta}^{\rm diag}(N)\left(\epsilon^{\beta\zeta\tau}\left(\delta_{\tau\alpha}\langle\hat \sigma^{\eta}\rangle+i\epsilon^{\tau\alpha\tau^{\prime}}\langle\hat \sigma^{\eta}\hat \sigma^{\tau^{\prime}}\rangle\right)+\epsilon^{\beta\eta\tau}\left(\delta_{\tau\alpha}\langle\hat \sigma^{\zeta}\rangle+i\epsilon^{\tau\alpha\tau^{\prime}}\langle\hat \sigma^{\zeta}\hat \sigma^{\tau^{\prime}}\rangle\right)\right.\nonumber\\
	&\left.\qquad\qquad\qquad+\epsilon^{\zeta\alpha\tau}\left(\delta_{\beta\tau}\langle\hat \sigma^{\eta}\rangle+i\epsilon^{\beta\tau\tau^{\prime}}\langle\hat \sigma^{\eta}\hat \sigma^{\tau^{\prime}}\rangle\right)+\epsilon^{\eta\alpha\tau}\left(\delta_{\beta\tau}\langle\hat \sigma^{\zeta}\rangle+i\epsilon^{\beta\tau\tau^{\prime}}\langle\hat \sigma^{\zeta}\hat \sigma^{\tau^{\prime}}\rangle\right)\right)\nonumber\\
	&\quad+\frac{1}{N}\sum_{\alpha,\beta}\Gamma_{\alpha,\beta}^{\rm diag}(N)\left(\epsilon^{\beta\zeta\tau}\left(\delta_{\tau\eta}\langle\hat \sigma^{\alpha}\rangle+i\epsilon^{\tau\eta\tau^{\prime}}\left(\delta_{\tau^\prime\alpha}+i\epsilon^{\tau^\prime\alpha\tau^{\prime\prime}}\langle\hat \sigma^{\tau^{\prime\prime}}\rangle\right)\right)\right.\nonumber\\
	&\qquad\qquad\qquad\qquad\qquad\left.+\epsilon^{\beta\eta\tau}\left(\delta_{\zeta\tau}\langle\hat \sigma^{\alpha}\rangle+i\epsilon^{\zeta\tau\tau^{\prime}}\left(\delta_{\tau^\prime\alpha}+i\epsilon^{\tau^\prime\alpha\tau^{\prime\prime}}\langle\hat \sigma^{\tau^{\prime\prime}}\rangle\right)\right)\right.\nonumber\\
	&\qquad\qquad\qquad\qquad\qquad\left.+\epsilon^{\zeta\alpha\tau}\left(\delta_{\tau\eta}\langle\hat \sigma^{\beta}\rangle+i\epsilon^{\tau\eta\tau^{\prime}}\left(\delta_{\beta\tau^\prime}+i\epsilon^{\beta\tau^\prime\tau^{\prime\prime}}\langle\hat \sigma^{\tau^{\prime\prime}}\rangle\right)\right)\right.\nonumber\\
	&\qquad\qquad\qquad\qquad\qquad\left.+\epsilon^{\eta\alpha\tau}\left(\delta_{\zeta\tau}\langle\hat \sigma^{\beta}\rangle+i\epsilon^{\zeta\tau\tau^{\prime}}\left(\delta_{\beta\tau^\prime}+i\epsilon^{\beta\tau^\prime\tau^{\prime\prime}}\langle\hat \sigma^{\tau^{\prime\prime}}\rangle\right)\right)\right),\nonumber
\end{align}
where we define the coefficients,
\begin{equation}
	\Gamma_{\alpha,\beta}^{\rm diag}(N)=\frac{i}{N}\sum_{j=1}^N\gamma_{\alpha,\beta}^{j,j},\quad\Gamma_{\alpha,\beta}^{\rm off}(N)=\frac{i}{N}\sum_{j=1}^N\sum_{k\neq j}\gamma_{\alpha,\beta}^{j,k}.
\end{equation}
In order for the dynamics to be scalable, we consider the above parameters  finite and well behaved in the thermodynamic limit. Specifically, 
\begin{equation}
	\lim_{N \rightarrow \infty} \Gamma_{\alpha, \beta}^{\rm off}(N) = \tilde{\Gamma}_{\alpha,\beta}^{\rm off}, \qquad 
	\lim_{N \rightarrow \infty} \Gamma_{\alpha, \beta}^{\rm diag}(N) = \tilde{\Gamma}_{\alpha,\beta}^{\rm diag},
\end{equation}
with $\tilde{\Gamma}_{\alpha,\beta}^{\rm off(diag)}$ finite.  Noticing that  $\lim_{N\rightarrow\infty}\langle\hat \sigma^{\alpha}\hat\sigma^{\beta}\cdots\rangle=\langle\hat m_{\alpha}\hat m_{\beta}\cdots\rangle + \lim_{N\rightarrow\infty}\mathcal{O}(1/N)$, we  can compute dynamic equations in the thermodynamic limit,
\begin{align}
	\frac{d}{dt} \langle\hat m_\zeta\rangle  
	&=\sum_{\alpha,\beta}\tilde{\Gamma}_{\alpha,\beta}^{\rm off}\left( \epsilon^{\beta\zeta\tau}\langle \hat m_{\tau}\hat m_{\alpha}\rangle
	+ \epsilon^{\zeta\alpha\tau}\langle \hat m_{\beta} \hat m_{\tau}\rangle\right)\nonumber\\
	&\quad+\sum_{\alpha,\beta}\tilde{\Gamma}_{\alpha,\beta}^{\rm diag} 
	\left( \epsilon^{\beta\zeta\tau}\left(\delta_{\tau\alpha}+i\epsilon^{\tau\alpha\tau^{\prime}}\langle \hat m_{\tau^{\prime}}\rangle\right)
	+ \epsilon^{\zeta\alpha\tau}\left(\delta_{\beta\tau}+i\epsilon^{\beta\tau\tau^{\prime}}\langle \hat m_{\tau^{\prime}}\rangle\right)\right),\nonumber\\
	\frac{d}{dt} \langle\hat m_\zeta\hat m_\eta\rangle 
	&=\sum_{\alpha,\beta}\tilde{\Gamma}_{\alpha,\beta}^{\rm off}\left(\epsilon^{\beta\zeta\tau}\langle\hat m_{\tau}\hat m_{\eta}\hat m_{\alpha}\rangle+\epsilon^{\beta\eta\tau}\langle\hat m_{\zeta}\hat m_{\tau}\hat m_{\alpha}\rangle+\epsilon^{\zeta\alpha\tau}\langle\hat m_{\beta}\hat m_{\tau}\hat m_{\eta}\rangle+\epsilon^{\eta\alpha\tau}\langle\hat m_{\beta}\hat m_{\zeta}\hat m_{\tau}\rangle\right)\\
	&\quad+\sum_{\alpha,\beta}\tilde{\Gamma}_{\alpha,\beta}^{\rm diag}\left(\epsilon^{\beta\zeta\tau}\left(\delta_{\tau\alpha}\langle\hat m_{\eta}\rangle+i\epsilon^{\tau\alpha\tau^{\prime}}\langle\hat m_{\eta}\hat m_{\tau^{\prime}}\rangle\right)+\epsilon^{\beta\eta\tau}\left(\delta_{\tau\alpha}\langle\hat m_{\zeta}\rangle+i\epsilon^{\tau\alpha\tau^{\prime}}\langle\hat m_{\zeta}\hat m_{\tau^{\prime}}\rangle\right)\right.\nonumber\\
	&\left.\qquad\qquad\qquad+\epsilon^{\zeta\alpha\tau}\left(\delta_{\beta\tau}\langle\hat m_{\eta}\rangle+i\epsilon^{\beta\tau\tau^{\prime}}\langle\hat m_{\eta}\hat m_{\tau^{\prime}}\rangle\right)+\epsilon^{\eta\alpha\tau}\left(\delta_{\beta\tau}\langle\hat m_{\zeta}\rangle+i\epsilon^{\beta\tau\tau^{\prime}}\langle\hat m_{\zeta}\hat m_{\tau^{\prime}}\rangle\right)\right).\nonumber
\end{align}

Deriving the Heisenberg equations of motion for the second cumulant in the thermodynamic limit, we obtain
\begin{align}
	\frac{d}{dt} K\left(\hat m_\zeta,\hat m_\eta\right)
	&=\sum_{\alpha,\beta}\tilde{\Gamma}_{\alpha,\beta}^{\rm off}\left(\epsilon^{\beta\zeta\tau}\left(K\left(\hat m_{\tau},\hat m_{\eta},\hat m_{\alpha}\right)+K\left(\hat m_{\tau},\hat m_{\eta}\right)K\left(\hat m_{\alpha}\right)+K\left(\hat m_{\eta},\hat m_{\alpha}\right)K\left(\hat m_{\tau}\right)\right)\right.\nonumber\\
	&\left.\qquad\qquad+\epsilon^{\zeta\alpha\tau}\left(K\left(\hat m_{\beta},\hat m_{\tau},\hat m_{\eta}\right)+K\left(\hat m_{\beta},\hat m_{\eta}\right)K\left(\hat m_{\tau}\right)+K\left(\hat m_{\tau},\hat m_{\eta}\right)K\left(\hat m_{\beta}\right)\right)\right.\nonumber\\
	&\left.\qquad\qquad+\epsilon^{\beta\eta\tau}\left(K\left(\hat m_{\zeta},\hat m_{\tau},\hat m_{\alpha}\right)+K\left(\hat m_{\zeta},\hat m_{\tau}\right)K\left(\hat m_{\alpha}\right)+K\left(\hat m_{\zeta},\hat m_{\alpha}\right)K\left(\hat m_{\tau}\right)\right)\right.\nonumber\\
	&\left.\qquad\qquad+\epsilon^{\eta\alpha\tau}\left(K\left(\hat m_{\beta},\hat m_{\zeta},\hat m_{\tau}\right)+K\left(\hat m_{\beta},\hat m_{\zeta}\right)K\left(\hat m_{\tau}\right)+K\left(\hat m_{\zeta},\hat m_{\tau}\right)K\left(\hat m_{\beta}\right)\right)\right)\\
	&\quad+i\sum_{\alpha,\beta}\tilde{\Gamma}_{\alpha,\beta}^{\rm diag}\left(\epsilon^{\beta\zeta\tau}\epsilon^{\tau\alpha\tau^{\prime}}K\left(\hat m_{\eta},\hat m_{\tau^{\prime}}\right)+\epsilon^{\zeta\alpha\tau}\epsilon^{\beta\tau\tau^{\prime}}K\left(\hat m_{\eta},\hat m_{\tau^{\prime}}\right)\right.\nonumber\\
	&\left.\qquad\qquad\qquad+\epsilon^{\beta\eta\tau}\epsilon^{\tau\alpha\tau^{\prime}}K\left(\hat m_{\zeta},\hat m_{\tau^{\prime}}\right)+\epsilon^{\eta\alpha\tau}\epsilon^{\beta\tau\tau^{\prime}}K\left(\hat m_{\zeta},\hat m_{\tau^{\prime}}\right)\right).\nonumber
\end{align}

We observe that $\overset{.}{K}(\hat m_{\zeta},\hat m_\eta) = f( h K(\hat m_\alpha) K(\hat m_\beta,\hat m_\gamma), h K(\hat m_\alpha,\hat m_\beta, \hat m_\gamma)) $ with $f$ a nonlinear function of its arguments and $h$ the possible coupling parameters $\Gamma_{[...]}^{[...]}$.
Therefore the same arguments of the collective case follows also here. E.g, assuming $h$ finite, given an initial uncorrelated  state (as \textit{e.g.} a product state) with 
\begin{equation}
	\label{eq.uncorrelatedMF.state.si}
	\lim_{N \rightarrow \infty}  K(\hat m_\alpha,\hat m_\beta) = 0, \, \lim_{N \rightarrow \infty}  K(\hat m_\alpha,\hat m_\beta,\hat m_\gamma) = 0,
\end{equation}
there is no generation of correlations $\overset{.}{K}(\hat m_\alpha,\hat m_\beta) = 0$ $\forall t$ and the dynamics is constrained to the mean filed first order cumulants, thus proving its exactness.

\section*{Correlated Initial States}

In this section we show evidences that gapless excitations and a diverging relaxation lifetime for the dynamics are also observed for different initial states, namely, correlated ones. Despite our analytical proof for gapless Lindbladian excitations is obtained for initial uncorrelated states, we expect it to have a broader validity. From one side 
the gapless condition assures the existence of gapless excitations at the Lindbladian superoperator level, i.e., generalized eigenoperators with a diverging lifetime (vanishing eigenvalue), which apart from systems with very specific
symmetries their action is not constrained to particular subspaces (e.g. uncorrelated states) within the full Hilbert space. 
Moreover, dissipative TC phases arise from the interactions among the subsystems, resulting in many-body
synchronized dynamics and a nonequilibrium phase robust to  imperfections. The initial condition has no major role on its support, rather  it is how the subsystems interact among themselves the important ingredient for their collective motion.

\begin{figure}
	\includegraphics[scale = 0.6]{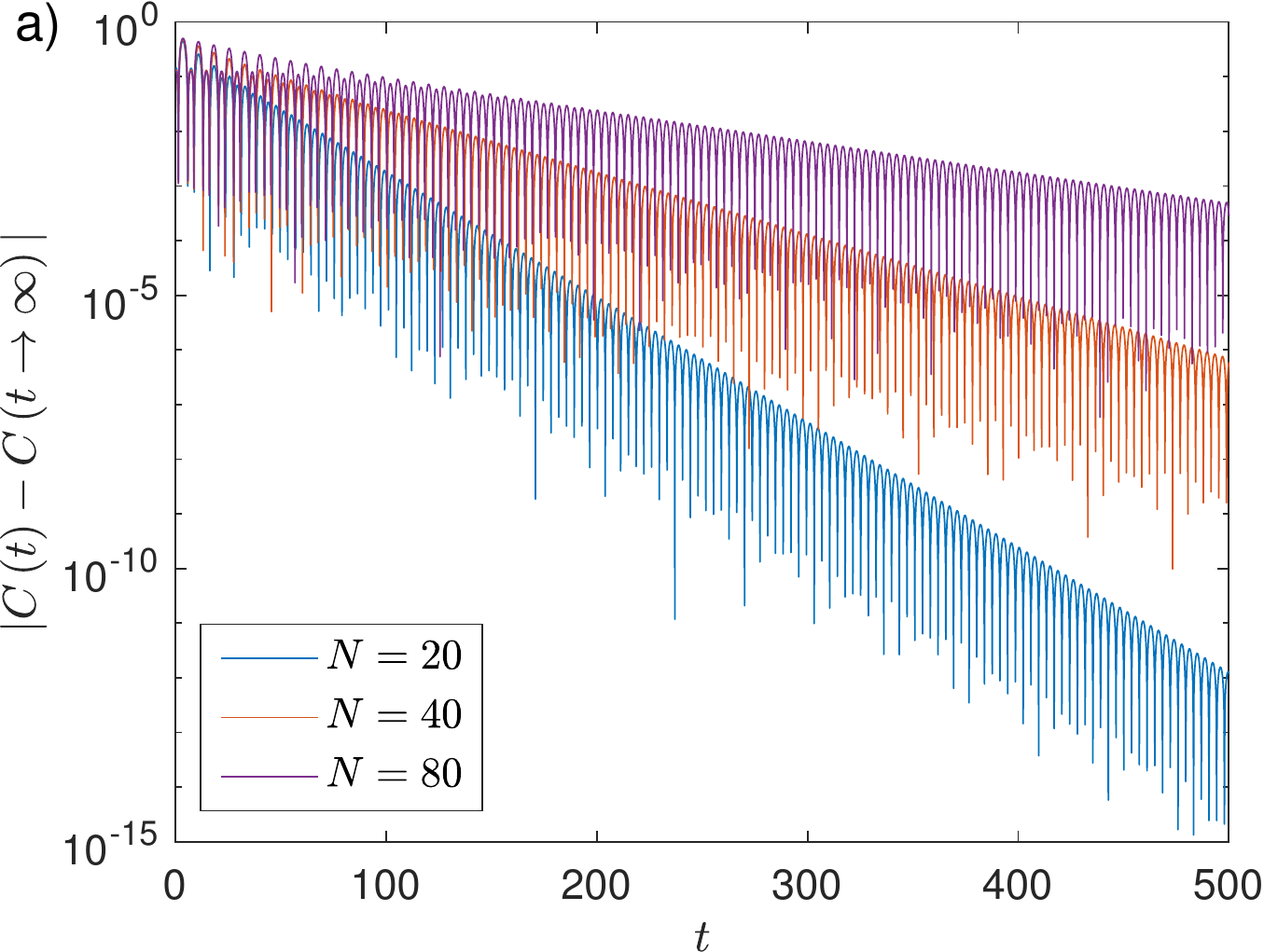}
	\includegraphics[scale = 0.6]{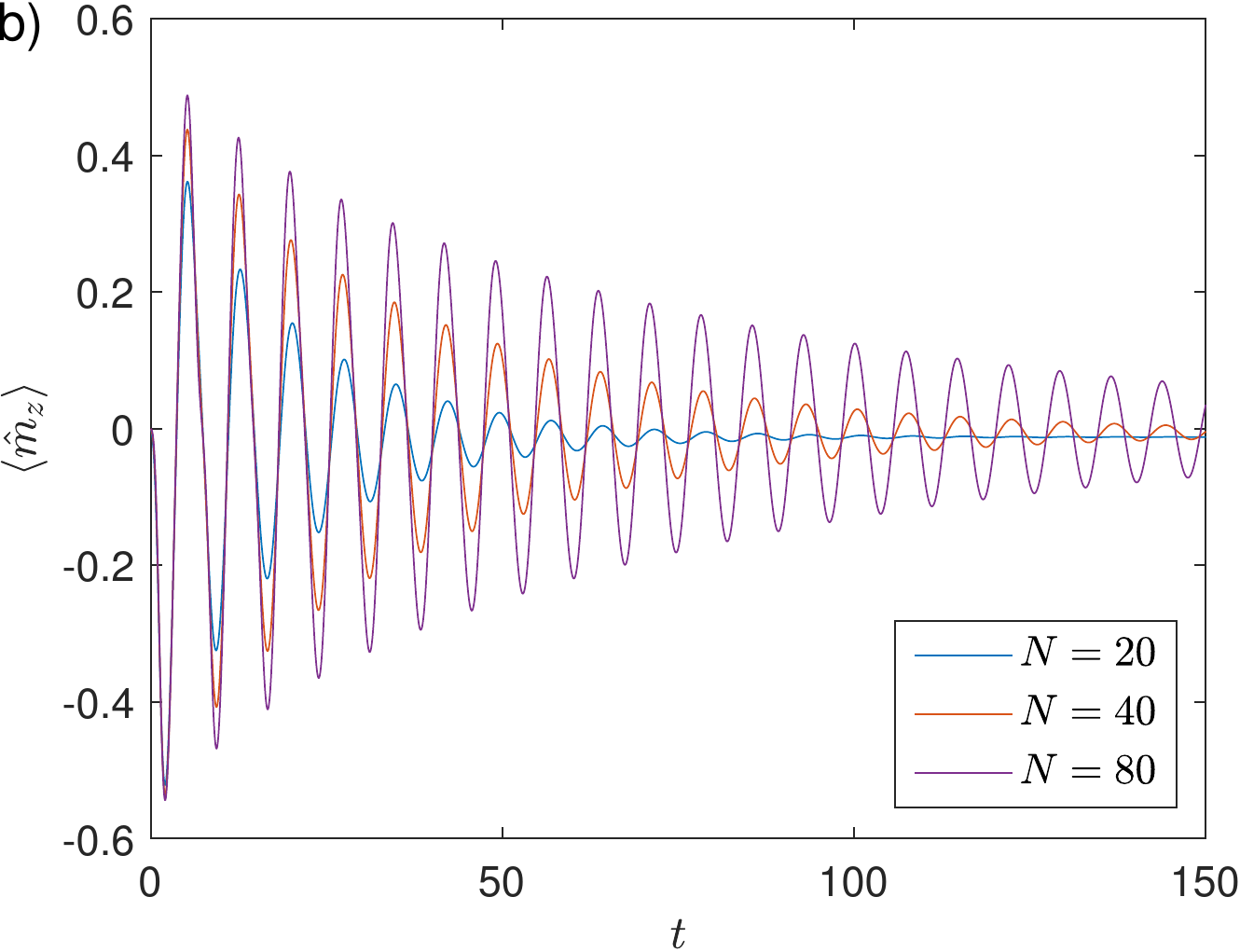}
	\caption{\textbf{(a)} Dynamics in log-lin scale of cumulants correlations $C=\sum_{\alpha} K(\hat m_\alpha,\hat m_\alpha)$ approaching its steady state value, in a collective spin-$1/2$ system for different system sizes. \textbf{(b)} Dynamics for the macroscopic magnetization along z-direction. In all cases we set $\omega_0/\kappa=2$ and the initial state is $\ket{\Psi^{+}}=\left(\ket{\uparrow\uparrow\cdots\uparrow}+\ket{\downarrow\downarrow\cdots\downarrow}\right)/\sqrt{2}$.}
	\label{fig:dyn_cum_su2}
\end{figure}

	In order to corroborate such arguments we have performed the dynamics for initial correlated states in the model of Eq.\eqref{eq:dissipation.su2}, where we show the presence of the gapless excitations and the persistent oscillations. We consider a maximally entangled initial state $\ket{\Psi^{+}}=\left(\ket{\uparrow\uparrow\cdots\uparrow}+\ket{\downarrow\downarrow\cdots\downarrow}\right)/\sqrt{2}$.  In Fig.(\ref{fig:dyn_cum_su2})-(left panel) we show the asymptotic dynamics for the cumulant correlations $C=\sum_{\alpha} K(\hat m_\alpha,\hat m_\alpha)$. We see it follows an exponential decay towards the steady state value, whose lifetime increases with system size, thus highlighting its gapless nature. In Fig.(\ref{fig:dyn_cum_su2})-(right panel) we show the dynamics for the macroscopic magnetization, whose oscillations lifetime diverge with system size leading to dissipative TC's.
	
\begin{thebibliography}{99}
\bibitem{footnote0} In its most general form, one could alternatively consider the $\{S^{\alpha}\}$ operators as an abstract SU(2) group, with $N$ representing the dimension of the irredutible representation for the algebra. Our discussion for the gapless sufficient condition follows similarly in this case.
\bibitem{footnote1} Despite our analytical proof is performed for an initial uncorrelated initial state, we believe it has a wider range for general initial states, as discussed in \cite{SM}.
\bibitem{footnote2} The dynamics for the density matrix with finite $N$ spins and mean-field equations were computed using Runge-Kutta numerical integration. We obtain the Lindbladian spectrum with numerical exact diagonalization of the superoperator matrix.
\end{thebibliography}
\end{document}